\documentclass[lineno]{jfm}

\usepackage{graphicx}
\usepackage{newtxtext}
\usepackage{newtxmath}
\usepackage{natbib}
\usepackage{hyperref}
\hypersetup{
    colorlinks = true,
    urlcolor   = blue,
    citecolor  = black,
}

\newcommand{\RomanNumeralCaps}[1]
\linenumbers

\usepackage{amsmath}
\usepackage{cases}
\usepackage{bm}
\usepackage{amssymb}
\usepackage{float}
\usepackage{caption} 
\usepackage{wrapfig}
\usepackage{cleveref}
\usepackage{xcolor}
\usepackage[labelformat=simple]{subcaption}		
\usepackage{tikz}
\usepackage{pst-node,pst-plot}

\usepackage{etoolbox}
\usepackage{pgfplots}
\usetikzlibrary{positioning}
\usetikzlibrary{calc,patterns,decorations.pathmorphing,decorations.markings}

\pgfplotsset{compat=newest}
\usetikzlibrary{plotmarks}
\usetikzlibrary{arrows.meta}
\usepgfplotslibrary{patchplots}
\usepackage{color}
\usetikzlibrary{shapes}
\usepackage{grffile}
\let\vec\bm
\usetikzlibrary{automata}
\usepgfplotslibrary{fillbetween}

\newcommand{\be}[1]{ \begin{equation} \label{eq:#1}}
\newcommand{\ee}{\end{equation}}

\newcommand{\bes}[1]{ \begin{equation*} \label{eq:#1}\begin{array}{rl}}
\newcommand{\ees}{\end{array}\end{equation*}}
\newcommand{\bess}[1]{ \begin{equation} \label{eq:#1}\begin{array}{rll}}
\newcommand{\eess}{\end{array}\end{equation}}

\newcommand{\besp}{ \begin{split} }
\newcommand{\eesp}{\end{split}}

\definecolor{burntorange}{RGB}{204, 85, 0}
\definecolor{red2}{rgb}{0.95000,0.40000,0.40000}%
\definecolor{red3}{rgb}{0.85000,0.20000,0.20000}%
\definecolor{red4}{rgb}{0.75000,0.10000,0.10000}%
\definecolor{blue2}{rgb}{0.50000,0.75000,1.00000}%

\DeclareRobustCommand\fullorange  {\tikz[baseline=-0.6ex]\draw[ultra thick,color=burntorange] (0,0)--(0.5,0);}
\DeclareRobustCommand\fullred  {\tikz[baseline=-0.6ex]\draw[ultra thick,color=red] (0,0)--(0.5,0);}

\DeclareRobustCommand\fullgray  {\tikz[baseline=-0.6ex]\draw[ultra thick,color=gray] (0,0)--(0.5,0);}
\DeclareRobustCommand\fullblue  {\tikz[baseline=-0.6ex]\draw[ultra thick,color=blue] (0,0)--(0.5,0);}
\DeclareRobustCommand\fullblack  {\tikz[baseline=-0.6ex]\draw[ultra thick,color=black] (0,0)--(0.5,0);}

\newrobustcmd*{\fullcustom}[1]{\tikz[baseline=-0.6ex]{\draw[ultra thick,color=#1] (0,0)--(0.5,0);}}
\newrobustcmd*{\fullcustomdashdotted}[1]{\tikz[baseline=-0.6ex]{\draw[ultra thick,color=#1, my dashdotted] (0,0)--(0.5,0);}}

\DeclareRobustCommand\arrowTriangle  {\tikz[baseline=-0.6ex]\draw[thick,-{Triangle[open]},color=black] (0,0)--(0.5,0);}
\DeclareRobustCommand\arrowCircle  {\tikz[baseline=-0.6ex]\draw[thick,-{Circle[open]},color=black] (0,0)--(0.5,0);}

\definecolor{mycolor4}{rgb}{0.49400,0.18400,0.55600}%
\definecolor{mycolor5}{rgb}{0.46600,0.67400,0.18800}%
\DeclareRobustCommand\fullgreen  {\tikz[baseline=-0.6ex]\draw[ultra thick,color=mycolor5] (0,0)--(0.5,0);}
\DeclareRobustCommand\fullpurple  {\tikz[baseline=-0.6ex]\draw[ultra thick,color=mycolor4] (0,0)--(0.5,0);}

\DeclareRobustCommand\dashedorange  {\tikz[baseline=-0.6ex]\draw[ultra thick,color=burntorange, my loosely dashed] (0,0)--(0.5,0);}
\DeclareRobustCommand\dashedgray  {\tikz[baseline=-0.6ex]\draw[ultra thick,color=gray, my loosely dashed] (0,0)--(0.5,0);}
\DeclareRobustCommand\dashedblack  {\tikz[baseline=-0.6ex]\draw[ultra thick,color=black, my loosely dashed] (0,0)--(0.5,0);}
\DeclareRobustCommand\dottedgray  {\tikz[baseline=-0.6ex]\draw[ultra thick,color=gray, loosely dotted] (0,0)--(0.5,0);}
\DeclareRobustCommand\dashdottededgray  {\tikz[baseline=-0.6ex]\draw[ultra thick,color=gray, my dashdotted] (0,0)--(0.5,0);}

\newlength{\radius}
\newrobustcmd*{\myfilledtriangle}[1]{\tikz{\draw[ultra thick, fill=#1] (90:\radius) -- (210:\radius) -- (330:\radius) -- cycle;}}
\newrobustcmd*{\myfilledcircle}[1]{\tikz{\draw[ultra thick, color=#1, fill=#1] (0,0)  circle (\radius);}}

\newrobustcmd*{\bullseye}{  \tikz[   even odd rule,    line width=0pt  ]{\fill (0,0) circle [radius=0.5em] (0,0) circle [radius=0.4em]; \fill (0,0) circle [radius=0.1em];  }}
\newrobustcmd*{\bullseyemagenta}{  \tikz[color=magenta,   even odd rule,    line width=0pt  ]{\fill (0,0) circle [radius=0.5em] (0,0) circle [radius=0.4em]; \fill (0,0) circle [radius=0.1em];  }}
    
\newrobustcmd*{\mysquare}[1]{\tikz{\draw[ultra thick,color=#1] (45:\radius) -- (135:\radius) -- (225:\radius) -- (315:\radius) -- cycle;}}
\newrobustcmd*{\mycircle}[1]{\tikz{\draw[ultra thick, color=#1] (0,0)  circle (\radius);}}
\newrobustcmd*{\mydashedcircle}[1]{\tikz{\draw[thick, color=#1,dash pattern=on 1.1pt off 1pt] (0,0) circle (0.1cm);}}
\newrobustcmd*{\mytriangle}[1]{\tikz{\draw[thick, color=#1] (90:\radius) -- (210:\radius) -- (330:\radius) -- cycle;}}

\newrobustcmd*{\squarered}[1]{\tikz{\draw[thick,red] (45:\radius) -- (135:\radius) -- (225:\radius) -- (315:\radius) -- cycle;}}
\newrobustcmd*{\squareblack}[1]{\tikz{\draw[thick,black] (45:\radius) -- (135:\radius) -- (225:\radius) -- (315:\radius) -- cycle;}}
\newrobustcmd*{\squareblue}[1]{\tikz{\draw[thick,blue] (45:\radius) -- (135:\radius) -- (225:\radius) -- (315:\radius) -- cycle;}}
\newrobustcmd*{\squaregray}[1]{\tikz{\draw[ultra thick, darkgray] (45:\radius) -- (135:\radius) -- (225:\radius) -- (315:\radius) -- cycle;}}
\newrobustcmd*{\redcircle}{\tikz{\draw[ultra thick, color=red] (0,0)  circle (\radius);}}
\newrobustcmd*{\bluecircle}{\tikz{\draw[ultra thick, color=blue] (0,0)  circle (\radius);}}
\newrobustcmd*{\blackcircle}{\tikz{\draw[ultra thick, color=black] (0,0)  circle (\radius);}}
\tikzstyle{my loosely dashed}=[dash pattern=on 6pt off 3pt]
\tikzstyle{my dashdotted}=[dash pattern=on 6pt off 3pt on 2pt off 3pt]

\title{Stability prediction of vortex induced vibrations of multiple freely oscillating bodies}

\author{Théo Mouyen \aff{1,2},
  Javier Sierra\aff{3},
  David Fabre\aff{1}\corresp{\email{david.fabre@imft.fr}}
 \and Flavio Giannetti\aff{2}}

\affiliation{\aff{1} IMFT, Institut de M\'ecanique des fluides de Toulouse, CNRS, Toulouse 31400, France
\aff{2}Università degli Studi di Salerno - Via Giovanni Paolo II, 132, 84084 Fisciano, Italia
\aff{3} ONERA Toulouse BP74025, F-31055, Toulouse, France}

\begin{document} 
\nolinenumbers 

\begin{nolinenumbers}
\maketitle
\end{nolinenumbers}

\nolinenumbers 

\nolinenumbers 

\begin{abstract}
The vortex-induced vibration of multiple spring-mounted bodies free to move in the orthogonal direction of the flow is investigated. In a first step, we derive a Linear Arbitrary Lagrangian Eulerian (L-ALE) method to solve the fluid-structure linear problem as well as a forced problem where a harmonic motion of the bodies is imposed. We then propose a low computational-cost impedance-based criterion to predict the instability thresholds. A global stability analysis of the fluid-structure system is then performed for a tandem of cylinders and the instability thresholds obtained are found to be in perfect agreement with the predictions of the impedance-based criterion. An extensive parametric study is then performed for a tandem of cylinders and the effects of mass, damping and spacing between the bodies are investigated. Finally we also apply the impedance-based method to a three-body system to show its validity to a higher number of bodies.
\end{abstract}


\section{Introduction}
\label{sec:intro}

Fluid-induced vibrations (FIV) are of great interest to many fields of engineering. 
They are generally classified either as vortex-induced vibrations (VIV), wake-induced vibrations (WIV) or galloping.
In the field of engineering, we find two main design philosophies. On the one hand, the design of structures that prevent such vibrations to avoid damage is an obvious example. 
\citet{griffin1982some} reviewed studies on vortex-induced vibrations of marine risers and listed means to suppress such oscillations. 
On the other hand, we find the design of oscillating/deformable structures that are conceived to harvest energy. Here, one aims to optimise the motion of the submerged rigid body or deformation of the flexible structure in order to harvest the most energy. Some examples include submerged oscillating/deformable structures that are able to convert energy from marine currents and waves \citep[see review from]{bernitsas2016harvesting}.
The Wave Carpet project developed by \citet{alam2012nonlinear} aims to extract energy from waves using a deformable carpet. In contrast,
the VIVACE concept from \citet{bernitsas2008vivace} proposes to convert kinetic energy from marine currents to electricity using vortex-induced vibrations of multiple oscillating cylinders.
Energy harvesting strategies also find applications in microfluidics: for instance, \citet{lee2019vortex} proposed a MEMS energy harvester based on the oscillation of a cylinder mounted on a piezoelectric chip.
In this context, at low Reynolds numbers, they found that the efficiency of the device was greater when placed in a dense field of oscillating cylinders.

The canonical case of a single freely-oscillating cylinder has been extensively studied with heavy focus on the lock-in phenomenon \citep{williamson2004vortex}.
It is defined as a synchronisation between the frequency associated to transverse oscillations of the rigid body and that of the vortex shedding in the wake of the cylinder.
Outside of the lock-in regime, however, the frequency tends to the vortex shedding frequency of a fixed cylinder.
\citet{navrose2016lock} found that the lock-in phenomenon induces high amplitude vibrations of the cylinder. It has also been shown that a decrease in the reduced mass ratio between the density of the body and the fluid leads to a wider synchronisation regime. Note that most studies with a single cylinder considered a single degree of freedom (1DOF) corresponding to transverse motion. According to \cite{williamson2004vortex}, in-line motion, if structurally allowed (2DOF), does not change much the dynamics, and mostly turns the transverse oscillation into a figure-eight trajectory where the streamwise motion is induced by a nonlinear effect. On the other hand, pure in-line oscillations seem not to have been observed for a single cylinder. Such motion would be linked to symmetric vortex shedding, which is likely not observed.

A number of configurations involving multiple freely oscillating bodies have been explored.
Authors have first focused on wake-induced vibrations (WIV). \cite{king1976wake} first explored WIV of flexible cylinders (two-degree of freedom i.e. 2DOF) in tandem traversing a free surface, either rigidly connected or not, for spacings of $L/D=[1.25 - 7]$, where $D$ is the diameter of the cylinders and $L$ is the distance between the centres, at Reynolds numbers $Re=\frac{U_\infty D}{\nu}=[10^3 - 2\times10^4]$, where $U_\infty$ is the inlet velocity and $\nu$ the fluid's viscosity. 
They observed both transverse oscillations and in-line oscillations, but the latter were only reported at much larger values of the Reynolds number, in the range $[10^3 - 2 \cdot 10^4$].

\cite{bokaian1984wake} focused on the transversal WIV (1DOF) of the rear body by fixing the front one.
In the interval $Re = [2900,5900]$, they found that the vortex shedding behind the front cylinder is suppressed for spacings of $L/D \leq 2$.
Later studies explored in detail the WIV of a rear oscillating body in the 2DOF \citep[]{brika1999flow} and 1DOF configurations \citep[]{assi2006experimental,assi2010wake}.
\citet{assi2010wake} found that for high separation between bodies, the amplitude of the rear body is decreased and resembles a VIV amplitude.
\citet{assi2013role} (1DOF) developed the concept of wake stiffness in the galloping of cylinders placed in tandem.
The steady lift across the wake is defined as a restoring force towards the center line, acting as a fluid dynamic spring.
The Strouhal number associated with the wake stiffness was found to be constant with the Reynolds number.
\citet{mittal2001flow} numerically studied the tandem and staggered configurations with a 2DOF configuration for low Reynolds number ($Re=100$) in the wake interference regime ($L/D=5.5$). 
For this large spacing, the front body behaves like an isolated cylinder with trajectories resembling an eight shape.
Soft lock-in was observed and the vortex-shedding frequency of the bodies is detuned from the natural frequency.
The rear body displays trajectories in the shape of an eight or a tilted ovoid, whether it is placed in tandem or in staggered configuration.
\citet{papaioannou2008effect} used an Arbitrary Lagrangian-Eulerian (ALE) method to further explore the effect of spacing on the 2DOF tandem.
For a $Re = 160$ and reduced mass $m^*=10$, they explored spacings ($L/D = [2.5,3.5,5.0]$) corresponding to different flow regimes in the fixed tandem case \citep[]{zdravkovich1987effects}.
Small values of the spacing lead to stronger oscillations of the upstream cylinder over a wider reduced velocity range and shift the response curves to higher reduced velocities.
\citet{borazjani2009vortex} directly simulated a tandem of cylinders with 1DOF for a low reduced mass, $L/D=1.5$ and $Re=200$. 
For low values of the reduced velocity ($U^*=2 \pi U_{\infty} \sqrt{m_{c} / k}/ D$, $k$ and $m_c$ being the stiffness and cylinder's mass respectively), they found that the oscillation amplitudes are small and therefore outside of the lock-in region.
The front cylinder exhibits larger oscillation amplitudes than the rear one. The effect of an increase of the reduced velocity is to bring the cylinders' oscillations out of phase, thus increasing their amplitudes of motion.
At a critical reduced velocity, the cylinders continue to oscillate out of phase but the rear cylinder's amplitude becomes greater than the front one. In particular, the authors found a wider lock-in region than for an isolated cylinder.
Besides, a structure that would be outside of the lock-in region can be brought into it by placing it in tandem with a similar structure.

\citet{kim2009aflow} experimentally studied the VIV of the tandem configuration with 1DOF transversal to the fluid flow for several spacings ($L/D = 1.1 - 4.2$) at  $Re = 4365-74200$. 
Five distinct regimes were identified. Regime I ($1.1 < L/D < 1.2$) features negligible vibrations due to minimal fluctuating lift forces, while Regime II ($1.2 < L/D < 1.6$) exhibits strong vibrations, particularly in the upstream cylinder, for higher reduced velocities.
Regime III ($1.6 < L/D < 3.0$) shows significant vibrations of both bodies, with the upstream cylinder's response being influenced by the downstream cylinder.
In Regime IV ($3.0 < L/D < 3.7$), vibrations are again minimal; the downstream cylinder stabilises the wake.
Finally, Regime V ($L/D > 3.7$) displays higher vibrations in the downstream cylinder, attributed to periodic Karman vortex shedding. 
In a subsequent study, \citet{kim2009bflow} used tripping wires to suppress vortex-induced vibrations.
They found that placing the wires at an optimal position effectively suppressed vibrations in flow regimes I–IV by altering the shear layer behaviour and preventing vortex formation.

\citet{prasanth2009vortex, prasanth2009flow} numerically examined the free vibrations of two cylinders in the staggered and tandem configurations with 2DOF at  $Re=100$ for $m^*=10$,  $L/D =5.5$ and compared the dynamic responses to those of a single cylinder.
In the staggered configuration, the upstream cylinder behaves similarly to a single cylinder but with slightly higher oscillation amplitudes, while the downstream cylinder exhibits significantly larger transverse oscillations. 
Lock-in occurs over a range wider than for a single cylinder, with shared vortex shedding frequencies.
The downstream cylinder in the staggered case displays both an eight-shape and orbital motions, influenced by complex vortex interactions and asymmetrical flow patterns.
For the tandem configuration, the upstream cylinder shows early lock-in and significant influence from the downstream cylinder, despite having a qualitatively similar transverse response to an isolated one.
The downstream cylinder experiences much larger oscillations that are twice that of a single cylinder in the laminar regime.
Its behaviour mimics high Reynolds number responses, including the presence of an upper branch in vibration response.
Both cylinders undergo synchronisation, with frequency and phase shifts tied to vortex shedding and lift forces.
Phase differences and hysteresis effects are observed, and the flow regime is divided into different regions based on flow-structure interactions.

\citet{griffith2017flow} investigated the dynamic response of staggered cylinders at $Re=200$ with 1DOF, for a fixed stream-wise spacing ($L/D = 1.5$).
They found that gap flow, which reverses direction as the cylinders oscillate, plays a critical role.
A regime map was developed, categorising major vortex shedding modes and temporal behaviours.
Unlike a single cylinder, matched natural and shedding frequencies do not produce synchronised oscillations; instead, quasi-periodic and chaotic responses emerge.
For rigid cylinders, three base modes were observed: no gap flow, gap pair dominated, and wake pair dominated—shifting with cross-stream offset.
Near the gap/wake pair transition, more complex flow states appear. When cylinders are free to oscillate, low reduced velocities yield minimal motion and rear-cylinder vortex shedding.
At intermediate velocities, out-of-phase oscillations enlarge the gap and produce an irregular vortex street.
At higher velocities, the rear cylinder chases the front, with joint vortex shedding. As the spacing increases, vortex pairs dominate and the system approaches single-cylinder behaviour.
\citet{huera2011vortex} conducted an experimental study of VIV and WIV of a tandem of flexible cylinders (2DOF) in the wake interference regime.
They found that both flexible cylinders in a tandem arrangement exhibit classical VIV near lock-in reduced velocities, regardless of the gap distance.
At higher reduced velocities, their dynamic responses diverge depending on the spacing between the bodies.
The upstream cylinder shows stronger VIV for smaller gaps, while the downstream cylinder may experience WIV at larger gaps, from the presence of vortex shedding in the gap region.
\citet{zhang2024flow} investigated fluid-induced vibrations (1DOF) of two square cylinders in tandem through simulations and reduced-order modelling.
Multiple vibration branches, such as VIV, biased oscillation, and galloping, are identified depending on reduced velocity and spacing ratio and their link to wake and structural modes is analysed. \citet{tirri2023linear} and \citet{zhang2024global} conducted a Linear Stability Analysis (LSA) at low Reynolds number of the tandem configuration with 1DOF over a very limited range of structural parameters.
They both found two leading unstable eigenmodes, one being associated to the classical vortex shedding behaviour in the wake of a tandem of fixed bodies and the other being of structural nature since they don't have a counterpart in the case of fixed bodies.

Some other authors explored configurations with more bodies and outlined the complexity of such systems. \citet{hosseini2021flow} have investigated the tandem configuration of both two and three fixed cylinders at $Re=200$. They found that when the distance between the two bodies is large, the wake exhibits a two-row vortex structure. Adding a third body in that wake has no impact in the majority of cases. However, a region was identified where the placement of an additional body suppresses the vortex shedding in the gap between the two upstream bodies.
The tandem of three cylinders oscillating transversely at low Reynolds number was also investigated by \citet{chen2018vortex,zhu2024fluid}. 

Note that, in most of the cited bibliography, configurations in which the cylinders are allowed to move in two directions (2DOF) essentially lead to the same dynamics as those in which the cylinder are only allowed to vibrate transversely (1DOF), driven by the transverse force due to antisymmetric vortex shedding. Pure in-line oscillation, which would be linked to symmetric vortex shedding, have not been reported, with the notable exception of  \cite{king1976wake}. However, such behaviour was only observed for $Re>10^3$, away from the range considered here. Likewise, no known study seem to have reported coupling with the third degree of freedom corresponding to rotational motion of the cylinder. Such motion would be linked to a torque, which is not expected to be significant for a cylinder, as it would be exerted by viscous stress only and not pressure, unlike for more elongated bodies where this kind of motion, known as flutter, is an important subject of research \citep{chai2021aeroelastic}.

The consequent number of parameters in a multi-body system renders exhaustive parametric studies time-consuming, which justifies the search for new methodologies that enable a systematic scanning of the problem's parameters. 
A strategy for analysing fluid-structure interactions has been to project the fluid forces onto the structural degrees of freedom and treat the problem as a harmonically forced structural oscillator, using generalized aerodynamic forces (GAF) to represent the fluid loading. This line of thought was formalised early on by \citet{hassig1971approximate}, who introduced the $p-k$ method later also referred as Schur complement formulation. In the original formulation, the unsteady aerodynamic forces are available for harmonic oscillations at discrete reduced frequencies. In practical implementations, the use of simplified aerodynamic models (analytical or semi-empirical), frequency interpolation for the approximation of the GAFs \citep{houtman2023global}, Kriging interpolation \citep{timme2011transonic} or POD-based \citep{bekemeyer2019flexible} has historically been a pragmatic way to reduce computational cost.

The idea of exploiting the unsteady forces exerted on a body undergoing prescribed oscillatory motion to predict instability was also pursued by \citet{sabino2020vortex}. In that work, the unsteady forces were obtained by the exact resolution of the linearized Navier–Stokes equations, rather than from  aerodynamic models or frequency interpolation. Subsequently, they defined a mechanical impedance as the ratio between the lift force and the velocity of the cylinders. This allowed to treat the problem by exploiting an analogy with electric systems \citep{conciauro1981meaning} for which an impedance of negative real part is a necessary condition for instability.
Note that a similar concept of impedance can be applied to acoustic systems, as was done by \citet{fabre2019acoustic} for an oscillating flow through a thin circular aperture.
It was then applied for the stability prediction of the flow  through a circular aperture in a thick plate \citep[see][]{fabre2020acoustic, sierra2022acoustic}

The objective of this study is to extensively explore the stability of the flow past a tandem of oscillating cylinders, exploiting the impedance method just outlined. As the coupling with the other degrees of freedom is not expected to be significant, we restrict to a single degree of freedom (1DOF) for each cylinder corresponding to transverse motion.
The instability of the tandem is first investigated through the fluid-structure problem in an Arbitrary Lagrangian-Eulerian frame and a Linear Stability Analysis is performed for several $(Re, m^*, U^*)$.
Secondly, we derive an impedance-based criterion from calculations of the forced case, i.e., where the cylinders are forced to oscillate sinusoidally in the transverse direction of the flow.
The results of the coupled case and the impedance-based predictions are compared and we then use the impedance-based method to explore the stability of the tandem for a vast range of parameters.

%
%
%

\section{Problem formulation }
\label{sec:}

We consider $N$ spring-mounted cylinders immersed in a Newtonian, incompressible fluid. 
Let $\tilde{\Omega}(t)$ denote the (deformable) fluid domain and  $\tilde{\Gamma}_i(t)$ its interface with the cylinders. The flow is physically parametrized by dynamic viscosity $\nu$, density $\rho$, incoming velocity $U_{\infty}$, diameter $D$ and spacing $L$ between the cylinders, yielding two dimensionless parameters, the Reynolds number $Re = \frac{U_\infty D}{\nu}$  and the spacing ratio $L/D$.
The fluid flow is described by the velocity and pressure fields $\tilde{\bm{u}}, \tilde{p}$, which are governed, in non-dimensional form, by the following set of equations and boundary conditions at the cylinder walls:

\begin{align}
\left.\frac{\partial \tilde{\bm{u}}}{\partial t}\right|_{\tilde{\bm{x}}}+(\tilde{\bm{\nabla}} \tilde{\bm{u}}) \cdot \tilde{\bm{u}}-\tilde{\bm{\nabla}} \cdot \tilde{\bm{\sigma}}(\tilde{\bm{u}}, \tilde{p})&=\bm{0} & \quad \text { in } \quad \tilde{\Omega}(t), \label{eq:momentum}\\ 
\tilde{\bm{\nabla}} \cdot \tilde{\bm{u}}&=0 &\quad \text { in } \quad \tilde{\Omega}(t), \label{eq:incomp} 
\\
 \tilde{\bf{u}}  & = \dot{Y}_i {\bf e}_y  & \quad \text { on } \quad \tilde{\Gamma}_i(t).
\label{eq:BC_cyl}
\end{align}
The symbol $(\tilde{\hphantom{h}})$  is used for time-dependent quantities as well as time and spacial derivatives evaluated in the time-dependent domain. $(\dot{\hphantom{h}})$ refers to time derivatives. The stress tensor is defined as $\tilde{\bm{\sigma}}(\tilde{\bm{u}}, \tilde{p})=-\tilde{p}  \bm{\text{I}} + 2\mu \tilde{\bm{D}}(\tilde{\bm{u}})$ with $\tilde{\bm{D}}(\tilde{\bm{u}}) = \left(\tilde{\bm{\nabla}} \tilde{\bm{u}}+\tilde{\bm{\nabla}} \tilde{\bm{u}}^{ {T}}\right)$. The notation $(|_{\tilde{\bm{x}}})$ refers to the time derivative in the deformed domain. This term contains the coupling with the motion of the body through the deformation of the domain, handled by the ALE framework as will be explained below. The coupling is also contained in the boundary condition \ref{eq:BC_cyl}. In the latter, $Y_i(t)$ is the instantaneous displacement of the $i$-th cylinder in the transverse direction.

The cylinders are physically parametrized by their masses $m_{c_i}$, spring stiffnesses $k_i$ and damping parameters $g_i$. In a nondimensional way, this yields three nondimensional parameters for each cylinder, namely a mass ratio $m_i^* = \frac{4 m_{c_i} }{\pi D^2 \rho_f}$,  a reduced velocity  $U^*_i = 2 \pi U_{\infty} \sqrt{m_{c_i} / k_i}/ D$, and a damping coefficient $\gamma_i = g_i/(2\sqrt{m_{c_i}k_i})$.
The equation governing the motion of the $i$-th cylinder is, in a non-dimensional form:

\begin{equation}
\ddot{Y}_i+\frac{4 \pi \gamma_i}{U_i^*} \dot{Y}_i+\left(\frac{2 \pi}{U_i^*}\right)^2 Y_i = \dfrac{4 F_{y_i}(t)}{\pi m_i^*},  \quad \text{ for }i=1,\dots, N, \label{eq:N_cyl}
\end{equation}
\begin{equation}
\mbox{ with } F_{y_i} =  \int_{\tilde{\Gamma}_i} \bm{e}_{y} \cdot \tilde{\bm \sigma} \cdot \bm{n} {~d} \tilde{\Gamma}_i.
\label{eq:defFy}
\end{equation}



\begin{figure}
\centering
\includegraphics[scale=1, trim = 0cm 0cm 0cm 0cm, clip]{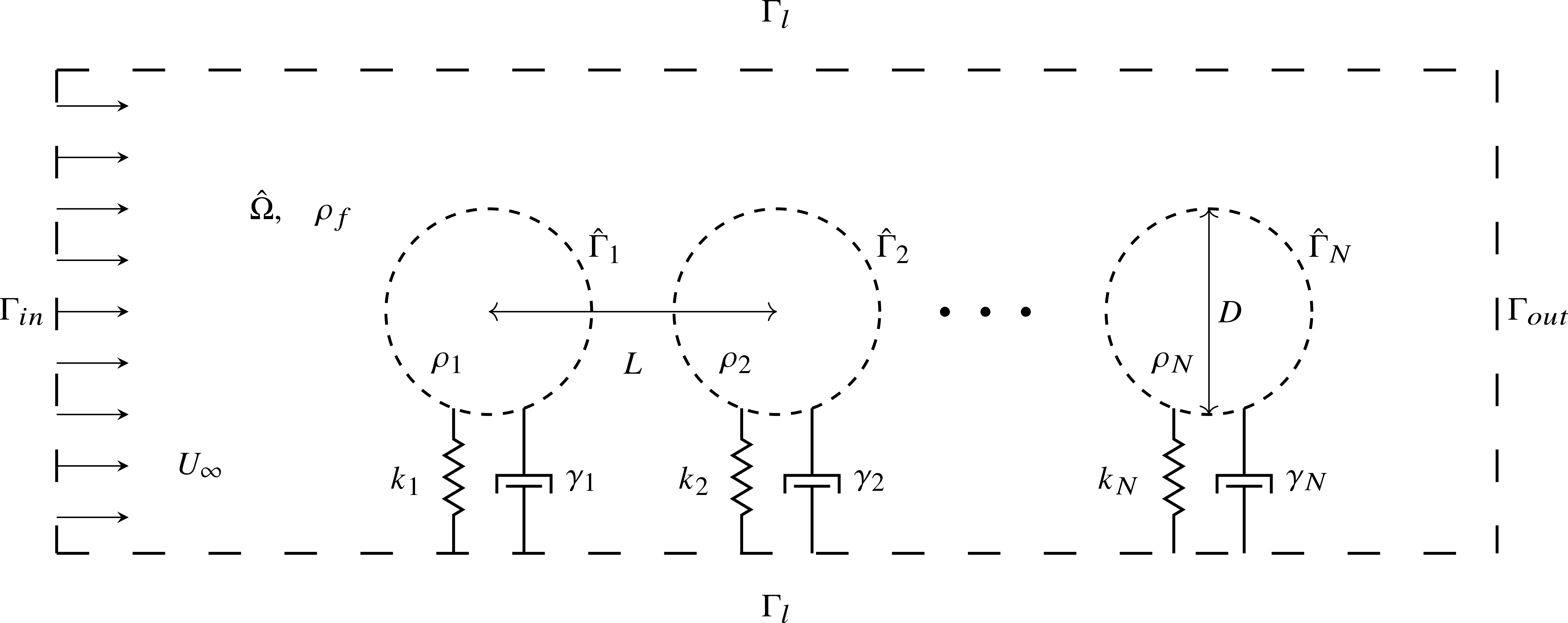}
\caption{Array of $N$ spring mounted cylinders with densities, spring stiffness and damping parameters: $\rho_i$, $k_i$ and $\gamma_i$. The cylinders are immersed in a fluid domain $\hat{\Omega}$ of density $\rho_f$. The domain is delimited by inlet and outlet boundaries, $\Gamma_{in}$ and $\Gamma_{out}$ as well as lateral boundaries $\Gamma_l$.}
\label{fig:CylinderSketch}
\end{figure}

\subsection {Arbitrary Lagrangian Eulerian formulation}

\subsubsection{General formalism}

\begin{figure}
\centering
\includegraphics[scale=1, trim = 0cm 0cm 0cm 0cm, clip]{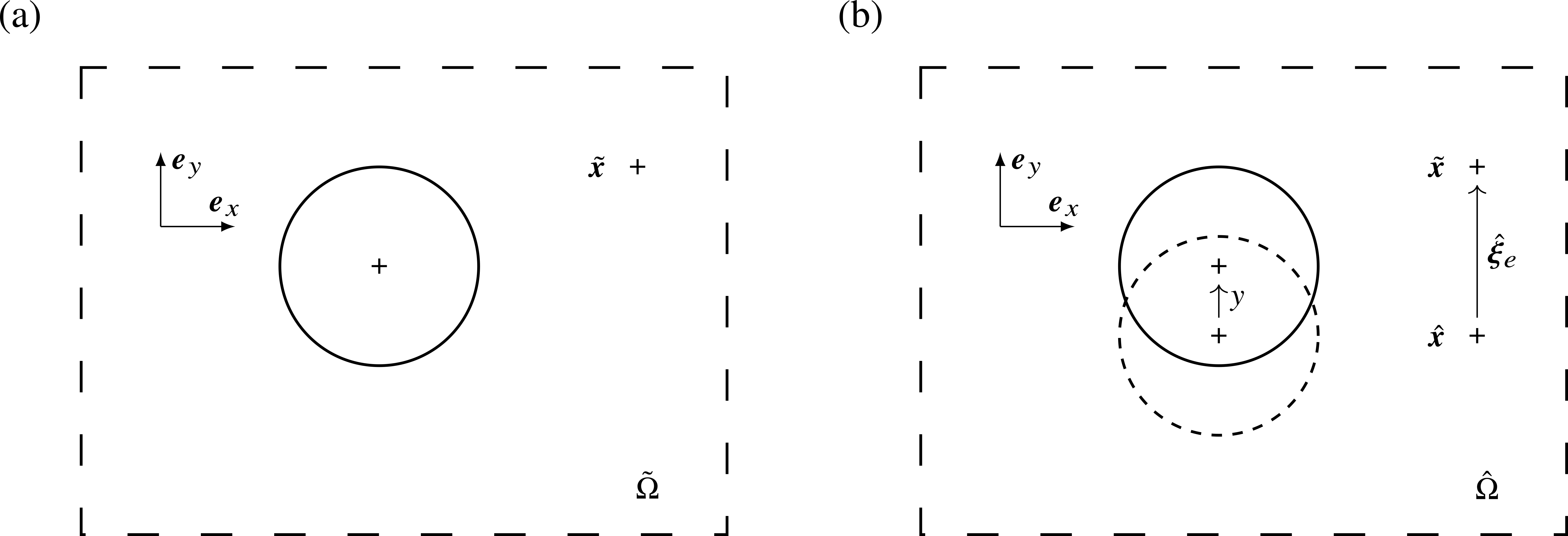}
\caption{Sketch of the geometrical transformations involved in the ALE approach: (a) Actual time-dependent domain and (b) reference domain.} 
\label{fig:ALE_diff}
\end{figure}

The Arbitrary Lagrangian Eulerian method is a conforming method that allows to treat interfaces in a Lagrangian frame of reference while the fluid is treated in an Eulerian frame of reference.
We consider a fixed reference domain $\hat{\Omega}$ where unknowns are evaluated in an Eulerian frame of reference.
Lagrangian variables on the other hand are evaluated on the actual physical domain $\tilde{\Omega}(t)$, which is time-dependent.
Let us define the diffeomorphism $\mathcal{A}$ that allows us to express the position $\tilde{\vec{x}}(t)$ of the actual domain with respect to the position $\hat{\vec{x}}$ of the reference domain:
\begin{equation}\label{eq:diffeo_ALE}
\begin{aligned}
\mathcal{A} : \hat{\Omega} \times \mathbb{R}^{+} &\longrightarrow \tilde{\Omega} \times \mathbb{R}^{+}, \\
(\hat{\bm{x}}, t) &\longmapsto\left(\tilde{\bm{x}}(t), t\right).
\end{aligned}
\end{equation}
This diffeomorphism allows a mapping of the actual domain through the position
\begin{equation}
    \tilde{\bm{x}} = \hat{\bm{x}}+\hat{\bm{\xi}}_e(\hat{\bm{x}}, t),
\end{equation}
where $\hat{\vec{\xi}}_e$ is an extension displacement field that propagates the Lagrangian interface deformation to the fluid domain (as schematised in figure \ref{fig:ALE_diff}).
This field is arbitrary and it is determined as a solution of an elliptic equation, $-\hat{\nabla} \cdot \bm{\mathcal{E}}\left(\hat{\bm{\xi}}_e\right)=\bm{0}$, which ensures a smooth distribution over the whole domain.
Following the methodology employed by \citet{pfister2019linear}, we apply the diffeomorphism to the Lagrangian variables and we substitute them into \cref{eq:momentum,eq:incomp} which yields the ALE formulation of the incompressible Navier-Stokes equation in a stress-free configuration:

\begin{align}
\hat{J}\left(\hat{\bm{\xi}}_e\right) \frac{\partial \hat{\bm{u}}}{\partial t}+\left( (\hat{\bm{\nabla}} \hat{\bm{u}}) \hat{\bm{\Phi}}\left(\hat{\bm{\xi}}_e\right)\right)\left(\hat{\bm{u}}-\frac{\partial \hat{\bm{\xi}}_e}{\partial t}\right)-\hat{\bm{\nabla}} \cdot \hat{\bm{\Sigma}}\left(\hat{\bm{u}}, \hat{p}, \hat{\bm{\xi}}_e\right)=\bm{0} \quad \text { in } \quad \hat{\Omega}, \label{eq:momentum_ALE}\\ 
-\hat{\nabla} \cdot \bm{\mathcal{E}}\left(\hat{\bm{\xi}}_e\right)=\bm{0} \quad \text { in } \quad \hat{\Omega}, \label{eq:ext_ALE} \\
-\hat{\bm{\nabla}} \cdot\left(\hat{\bm{\Phi}}\left(\hat{\bm{\xi}}_e\right) \hat{\bm{u}}\right)=0 \quad \text { in } \quad \hat{\Omega}.
\label{eq:incomp_ALE}
\end{align}

In the previous expressions, $\hat{\bm{\nabla}}$ is the gradient in the reference coordinates. $\hat{\bm{\Phi}}\left(\hat{\bm{\xi}}_e\right)=\hat{J}\left(\hat{\bm{\xi}}_e\right) \hat{\bm{F}}\left(\hat{\bm{\xi}}_e\right)^{-1}$ denotes the deformation operator introduced by the change of variables, with $\hat{J}\left(\hat{\bm{\xi}}_e\right)=\operatorname{det}\left(\hat{\bm{F}}\left(\hat{\bm{\xi}}_e\right)\right)$ the Jacobian of the deformation gradient $\hat{\bm{F}}\left(\hat{\bm{\xi}}_e\right)=\bm{\text{I}}+\hat{\bm{\nabla}} \hat{\bm{\xi}}_e$.
The ALE fluid stress tensor expressed in the reference configuration writes as 
\begin{equation}
\hat{\bm{\Sigma}}\left(\hat{\bm{u}}, \hat{p}, \hat{\bm{\xi}}_e\right)=\hat{\bm{\sigma}}\left(\hat{\bm{u}}, \hat{p}, \hat{\bm{\xi}}_e\right) \hat{\bm{\Phi}}\left(\hat{\bm{\xi}}_e\right)^{ {T}},
\label{StressTensorALE}
\end{equation}
where 
$\hat{\bm{\sigma}} =-\hat{p} \bm{\text{I}}+\frac{1}{Re} \hat{\bm{D}} $,  
with the viscous dissipation tensor defined as
\begin{equation}
\hat{\bm{D}}\left(\hat{\bm{u}}, \hat{\bm{\xi}}_e\right)=\frac{1}{2} \frac{1}{\hat{J}\left(\hat{\bm{\xi}}_e\right)}\left((\hat{\bm{\nabla}} \hat{\bm{u}}) \hat{\bm{\Phi}}\left(\hat{\bm{\xi}}_e\right)+\hat{\bm{\Phi}}\left(\hat{\bm{\xi}}_e\right)^{ {T}}(\hat{\bm{\nabla}} \hat{\bm{u}})^{ {T}}\right).
\label{ViscousStressTensorALE}
\end{equation}
Hereinafter, we particularize the elliptic operator $\bm{\mathcal{E}} = \nabla$, that is, the extension displacement field is determined by solving a Laplace equation. The complete formulation used to determine the extension field is as follows,
\begin{numcases}{} 
\Delta \hat{\bm{\xi}}_e = \bm{0}, \label{eq:LaplaceDisplacementField} \\
\hat{\bm{\xi}}_{e} = Y_i {\bf e}_y \quad \text{on} \quad \hat{\Gamma}_i. 
 \end{numcases}

\subsubsection{The discrete-ALE ansatz}

Equation \ref{eq:LaplaceDisplacementField} is linear, and the number of cylinders is finite; thus, according to the superposition principle, we can look for a solution of the extension field as a function of the cylinders' vertical displacement:

\begin{equation}
\hat{\bm{\xi}}_e = \sum_{i=1}^{N} Y_i(t) \bm{\xi}_{e_i},
\label{eq:Discrete ALE}
\end{equation}
where $\bm{\xi}_{e_i}$ is an elementary field associated to the displacement of the $i$-th cylinder, namely the solution of the elementary problem 
\begin{numcases}{} 
\Delta \bm{\xi}_{e_i} = \bm{0}, \\
\bm{\xi}_{e_i} = {\bf e}_y \quad \text{on} \quad \hat{\Gamma}_i, \\
\bm{\xi}_{e_i} = {\bf 0}  \quad \text{ on} \quad \hat{\Gamma}_{j, i \neq j} .
\label{eq:LaplaceDisplacementFieldElementary}
 \end{numcases}
We call (\ref{eq:Discrete ALE}) the {\em discrete-ALE ansatz}. The main advantage of this expression is that it is sufficient to solve $N$ elementary problems once to reconstruct the deformation field for all possible values of $Y_i$,  allowing for a reduction of the computational time required to solve the entire system.


\subsection {Linearized VIV problem}

Following the usual approach, both the fluid and structural variables are decomposed into a steady component and a small-amplitude perturbation which is searched under a modal form. Namely, for the fluid part of the problem, we start with the expansion 
 \begin{equation} \label{eq:perturbation}
    \hat{\bm{q}}_f(\hat{\bm{x}}, t)
    =\bm{q}_{f,0}(\hat{\bm{x}}) 
    + \varepsilon {\bm{q}}_f(\hat{\bm{x}}) 
    e^{-i \omega t}
\end{equation}
were  $\bm{q}_{f,0}$ is the so-called {\em base flow}, corresponding to the steady solution of the Navier-Stokes equations in the reference domain,  $\varepsilon \ll 1$, $\bm{q}_f$ is the fluid part of the eigenmode, and  $\omega$ is an a-priori complex eigenvalue. Unlike in other conventions, the perturbation are chosen to be written without caret symbol. They remain however quantities that are evaluated on the reference domain.

Similarly, for the structural part of the problem, we parametrize the displacement of the cylinders by
\begin{equation} \label{eq:perturbationY}
   Y_i(t) = \varepsilon y_i e^{-i \omega t} ; \qquad \dot{Y}_i(t) = \varepsilon z_i e^{-i \omega t} \quad \mbox{ with } z_i = - i \omega y_i. 
\end{equation}
The eigenmode of the  fluid-structure problem is thus defined as 
${\bm{q}} = \left({\bm{q}}_f, y_1, ... ,y_N,z_1, ..., z_N \right)$, 
where ${\bm{q}}_f=\left({\bm{u}},{p}\right)$ denotes its "fluid" part  
and $[y_1,...,y_N,z_1,..,z_N]$ its "solid part".


\subsubsection{ALE fluid-structure coupled formulation}

Substituting the ansatz \cref{eq:perturbation} into \cref{eq:momentum_ALE,eq:incomp_ALE,eq:N_cyl,eq:BC_cyl} 
and writing the equations in the steady deformed configuration \citep[see][for more details]{pfister2019linear} leads to the following formulations: 

\begin{equation} \label{eq:momentum_ALE_lin}
\begin{aligned}
&-i\omega\left(\underbrace{ {\bm{u}}}_{\vec{B}_{ff} {\bf q}_f } + \sum_{i=1}^{N} y_i (\underbrace{-\bm{\xi}_{e_i} \cdot \hat{\bm{\nabla}} \vec{u}_0}_{\vec{B}_{ui} y_i })\right) 
= \underbrace{- \left( {\bm{u}} \cdot \hat{\bm{\nabla}} \vec{u}_0  + \vec{u}_0 \cdot \hat{\bm{\nabla}} {\bm{u}}  \right) + 2\mu \hat{\bm{\nabla}} \cdot \hat{\bm{D}}\left({\bm{u}}\right)}_{\vec{A}_{uu} {\bf q}_f} \underbrace{-\hat{\bm{\nabla}} p}_{\vec{A}_{up} {\bf q}_f} \\
&
+ \sum_{i=1}^{N}  \Bigg[ \underbrace{- \vec{u}_0 \cdot \Big( \left(\hat{\bm{\nabla}} \vec{u}_0\right) \big( (\hat{\bm{\nabla}} \cdot \bm{\xi}_{e_i}) \bm{\text{I}} - \hat{\bm{\nabla}} \bm{\xi}_{e_i}\big) \Big) 
 - \hat{\bm{\nabla}} \cdot \Big[ p_0  \bm{\text{I}} \big( (\hat{\bm{\nabla}} \cdot \bm{\xi}_{e_i}) \bm{\text{I}} - \hat{\bm{\nabla}} \bm{\xi}_{e_i}\big)^T \Big] \Bigg] y_i }_{\vec{A}^{(1)}_{ui} y_i} \\
&
+ \sum_{i=1}^{N} \Bigg[ \underbrace{-\mu \hat{\bm{\nabla}} \cdot \left( \left(\hat{\bm{\nabla}} \vec{u}_0 \right)  \left(\hat{\bm{\nabla}} \bm{\xi}_{e_i} \right) + \left(\hat{\bm{\nabla}} \bm{\xi}_{e_i}\right)^T \left(\hat{\bm{\nabla}} \vec{u}_0 \right)^T\right)
- 2 \mu \hat{\bm{\nabla}} \cdot \Big( \tilde{\bm{D}}(\vec{u}_0) \big( (\hat{\bm{\nabla}} \cdot \bm{\xi}_{e_i}) \bm{\text{I}} - \hat{\bm{\nabla}} \bm{\xi}_{e_i}\big)^T \Big) \Bigg] y_i
}_{\vec{A}^{(2)}_{ui} y_i},
\end{aligned}
\end{equation}
\begin{equation}
0 = \underbrace{\hat{\bm{\nabla}} \cdot {\bm{u}}}_{\vec{A}_{pu} {\bf q}_f} 
+ \sum_{i=1}^{N}  \underbrace{ \hat{\bm{\nabla}} \cdot \Big(  \big( \hat{\bm{\nabla}} \cdot \bm{\xi}_{e_i} \bm{\text{I}} - \hat{\bm{\nabla}} \bm{\xi}_{e_i}\big)  \vec{u}_0 \Big) y_i}_{\vec{A}_{pi} y_i}.
\end{equation}
The boundary conditions on the object's surface are
\begin{numcases}{} 
{\bm{u}} \cdot \bm{e}_y = z_i \quad \text{on} \quad \hat{\Gamma}_i, \\
{\bm{u}} \cdot \bm{e}_y = 0  \quad \text{ on} \quad \hat{\Gamma}_{j, i \neq j} .
\label{eq:BC}
 \end{numcases}
They are symbolically noted as $-i\omega \sum \vec{B}_{fi}^* y_i + \vec{A}_{ff}^* {\bm{q}}_f=0$,
where  $\vec{B}_{fi}^*$ and $\vec{A}_{ff}^*$ are restriction operators extracting the degrees of freedom localized along the boundaries of the cylinders.

The linearization of \cref{eq:momentum_ALE,eq:incomp_ALE} thus introduces the operators $\vec{A}_{ff} = \vec{A}_{uu} + \vec{A}_{up} + \vec{A}_{pu} + \vec{A}_{ff}^*$ and $\vec{B}_{ff}$ that are purely driven by fluid variables, and the operators $\vec{A}_{fi}  = \vec{A}^{(1)}_{ui} + \vec{A}^{(2)}_{ui} + \vec{A}_{pi} $ and $\vec{B}_{fi} = \vec{B}_{ui} + \vec{B}_{fi}^* $ that arise from the interaction of fluid and ALE variables. In this way, \cref{eq:momentum_ALE,eq:incomp_ALE} can be symbolically written with the previously defined operators as 
\begin{equation}
-i\omega \left( \vec{B}_{ff} \vec{q}_f + \sum_{i=1}^{N} \vec{B}_{fi} y_i \right) = \vec{A}_{ff} \vec{q}_f +  \sum_{i=1}^{N}  \vec{A}_{fi}y_i,   \quad  \text{ in } \hat{\Omega}_{f}.
\label{FluidOperators}
\end{equation}

\subsubsection{Cylinder's equations}

The lift force $F_{y_i}$ acting on the $i$-th cylinder was defined previously in primitive coordinates by \ref{eq:defFy}. Using the ALE ansatz and the definition \ref{StressTensorALE}
of the stress tensor, one is led to an expression of the form
\begin{equation}
F_{y_i} = \vec{F}_{if} \vec{q}_f  + \sum_{j=1}^{N} {F}_{ij}^* y_j
\label{eq:Fyi}
\end{equation}
This expression is composed of two terms. The first is found by integrating on the boundary the stress which is purely linked to the fluid motion:
\begin{equation} \label{eq:lift}
\vec{F}_{if} \vec{q}_f = 
\int_{\hat{\Gamma}_i} \big(-p \bm{n}+2\mu{\tilde{\bm{D}}}({\bm{u}}) \bm{n}\big) \cdot \bm{e}_{y} {~d} \hat{\Gamma}_i.
\end{equation}
The second component contains the effect of the deformation of the domain associated with the ALE method, and thanks to the discrete-ALE ansatz, it depends only upon the elementary extension fields $\bm{\xi}_j$ associated with each of the cylinders:

\begin{eqnarray}
 {F}_{ij}^* =   
 \displaystyle \int_{\hat{\Gamma}_i}  &
 \big(-p_0 \bm{\text{I}} + 2\mu{\tilde{\bm{D}}}(\vec{u}_0)\big) \big( \hat{\bm{\nabla}} \cdot  \bm{\xi}_{e_j} \bm{\text{I}} - \hat{\bm{\nabla}} \bm{\xi}_{e_j} \big)^T   \bm{n} \cdot \bm{e}_{y} {~d} \hat{\Gamma}_i \nonumber \\ \vspace{2 mm}
- \displaystyle \int_{\hat{\Gamma}_i}  & \mu  \left( (\hat{\bm{\nabla}} \vec{u}_0) (\hat{\bm{\nabla}} \bm{\xi}_{e_j}) +  (\hat{\bm{\nabla}} \bm{\xi}_{e_j})^T (\hat{\bm{\nabla}} \vec{u}_0)^T \right) \bm{n} \cdot \bm{e}_{y} {~d} \hat{\Gamma}_i.
\label{eq:lift_ALE}
\end{eqnarray}

Introducing \cref{eq:perturbation} in \cref{eq:N_cyl}, we obtain the following system for $i=1,\dots, N$
\begin{align}
\label{eq:N_cyl_eig_1}
-i\omega y_i &= z_i, \\
-i\omega z_i &= - \dfrac{4 \pi \gamma_i}{U_i^*} z_i - \big( \dfrac{2 \pi}{U^\ast_i} \big)^2 y_i + 
\frac{4}{\pi m^*_i} \left(
\vec{F}_{i,f} \vec{q}_f  + \sum_{j=1}^{N} {F}_{i,j}^* y_j \right).
\end{align}

\subsubsection{Eigenvalue formulation for the coupled problem}
Considering the coupled problem formulated in terms of the state-vector $\bm{q}$ containing both the fluid part $\bm{q}_f$ and the solid part $[y_1,...y_N,z_1,..,z_N]$,
the equations detailed in the two previous subsections can be written in the following matricial system

\begin{equation} \label{eq:matricial_eq}
    -i\omega \vec{B} \bm{q} = \vec{A} \bm{q},
\end{equation}

with matrices 
\begin{equation}
    \vec{B} = \begin{bmatrix} 
 \vec{B}_{ff} 	& \vec{B}_{f1} 	& \dots 	& \vec{B}_{fN}  &   0  & \dots 	& 0   \\
        		& 1	   	&          	&       	&      	 &       	&      \\
        		&         	&\ddots	&       	&       &   (0) 	&      \\
        		&         	&          	& 1 		&       &       	&      \\
        		&         	&          	&      	 	& 1 	 &       	&      \\
        		&   (0)  	&          	&       	&       &\ddots 	&      	\\
        		&         	&          	&       	&       &       	& 1
    \end{bmatrix},
\end{equation}
and
\begin{equation}
    \vec{A} = \begin{bmatrix} 
    \vec{A}_{ff} 	& \vec{A}_{f1}   						& \dots 	& \vec{A}_{fN} 						& 0     					& \dots  	& 0      					\\
        		&       						&       	&       						& 1   						&        	&       					\\
        		&       						&  (0)  	&       						&       					& \ddots 	&       					\\
        		&       						&       	&       						&       					&        	& 1    					\\
  \frac{4} {\pi m^*} \vec{F}_{1f}   &  \frac{4} {\pi m^*} F_{11}^*-( \frac{2 \pi}{U^*_1})^2  	&        	&     \frac{4} {\pi m^*}  F_{1N}^*         				& - \frac{4 \pi \gamma_1}{U_1^*}&        	&            	 				\\
    \vdots 	& \vdots                     			& \ddots 	& \vdots                    				&           					& \ddots 	&             					\\
    \frac{4} {\pi m^*} \vec{F}_{1N}  &  \frac{4} {\pi m^*} F_{N1}^*           				&        	&  \frac{4} {\pi m^*} F_{NN}^*-( \frac{2 \pi}{U^*_N})^2 	&           					&       	& - \frac{4 \pi \gamma_N}{U_N^*}
    \end{bmatrix}.
\end{equation}

\subsection {Forced problem \& Impedance}

Besides the resolution of the coupled problem as an eigenvalue problem as just described, we will also make use of an alternative method which consists of first considering the {\em forced problem} in which the motion of the cylinders are imposed to behave harmonically, i.e. $Y_i(t) = y_i e^{-i\omega t}$ with imposed amplitudes $y_i$ and real frequency $\omega$. Thanks to the linearity of the problem \ref{FluidOperators}, we can express its solution in compact form as

\begin{equation} \label{eq:qfj}
\vec{q}_f = \sum_{j=1}^{N}  \vec{q}_{fj} y_j, \quad \mbox{with} \quad   \vec{q}_{fj} = - \left[ \vec{A}_{ff} + i\omega \vec{B}_{ff}  \right]^{-1} \left[ \vec{A}_{fj} + i\omega \vec{B}_{fj} \right].
\end{equation}
In practice, each ${\bf q}_{fj}$ is the solution of the forced problem \ref{eq:momentum_ALE_lin} considering a unitary displacement of the $j$-th cylinder, namely
\begin{equation} \label{eq:StructuralOperators_forced}
y_j = 1 ; \quad \quad z_j = -i\omega ; \quad y_{i\ne j} =z_{i\ne j} =0.
\end{equation}
One can now introduce the decomposition \ref{eq:qfj} into the definition \ref{eq:Fyi}, of the lift forces acting on the $i$-th body. Using the operators defined in \ref{eq:lift} and \ref{eq:lift_ALE}, this leads to
\be{Fyi2}
F_{y_i} = \sum F_{ij} y_j \quad \text{where} \quad F_{ij} = \vec{F}_{if} \vec{q}_{fj} + F_{ij}^*
\ee



Each of the terms $F_{ij}$ can be considered as a {\em transfer function }, corresponding to the ratio of the force exerted on the body $i$ to the displacement of the body $j$.
Note that each term $F_{ij}$ depends only upon the elementary displacement field ${\bf \xi}_{e_j}$ and the elementary solution ${\bf q}_{fj}$ of the forced problem, both 
 calculated by imposing $y_j = 1$, $z_j = -i\omega$, $y_{i \neq j} = 0$ \text{and} $z_{i \neq j} = 0$. In other words, we impose the movement of the $j$-th cylinder and fix all others in order to calculate $F_{ij}$.

Rather than this definition as a transfer function, it turns out to be more physically significant to define an {\em impedance} $Z_{ij}$ relating the force on body $i$ to the opposite of the velocity of body $j$, hence the definition
\begin{equation} Z_{ij} = - \frac{F_{ij}}{z_j} \equiv \frac{F_{ij}}{i \omega}. \end{equation}
Note that this definition is equivalent to the one used in \cite{sabino2020vortex}, except for a factor 2 due to the fact that they founded their definition upon the lift coefficient $C_y$ instead of the dimensionless force $F_y = C_y/2$.

In a compact form, the impedance $Z_{ij}$ can also be expressed in terms of the previously introduced operators as 
\begin{equation} \label{eq:defZij}
Z_{ij} = (i \omega)^{-1} \left\{ \vec{F}_{if} \left[ \vec{A}_{ff} + i\omega \vec{B}_{ff}  \right]^{-1} \left[ \vec{A}_{fj} + i\omega \vec{B}_{fj} \right] + F_{ij}^* \right\} .
\end{equation}

\subsection {Generalised impedance criterion for a tandem of cylinders}
\label{sec:imp}

We will now focus on the case of a tandem of cylinders ($N=2$). Solving the forced problem for the front and rear cylinder will respectively give the impedances $Z_{11}$, $Z_{21}$ and $Z_{12}$, $Z_{22}$.
We can plug equation \ref{eq:lift_ALE} along with the definition of the impedances into the harmonic oscillator equations \ref{eq:N_cyl_eig_1} of the  fluid-structure problem, which yields:

\begin{equation}
\begin{cases} \label{eq:imp_harmonic}
 \left(-\omega^2 - \dfrac{4\pi \gamma_1}{U_1^*} i \omega + \left( \dfrac{2\pi}{U_1^*} \right) ^2 \right)y_1 = \dfrac{i \omega}{\pi m_1^*} 4\left( Z_{11}y_1 + Z_{21}y_2 \right), \\[3ex]
 \left(-\omega^2 - \dfrac{4\pi \gamma_2}{U_2^*} i \omega + \left( \dfrac{2\pi}{U_2^*} \right) ^2 \right)y_2 = \dfrac{i \omega}{\pi m_2^*} 4\left( Z_{12}y_1 + Z_{22}y_2 \right). 
\end{cases}
\end{equation}
Building the matrix
\begin{equation} \label{eq:imp_matrix}
    Z_T = \begin{bmatrix} 
    - \omega^2 - \dfrac{4\pi \gamma_1}{U_1^*} i \omega + \left( \dfrac{2\pi}{U_1^*} \right) ^2 - \dfrac{i \omega 4 Z_{11}}{\pi m_1^*}  & - \dfrac{i \omega 4 Z_{21}}{\pi m_1^*}    \\
    - \dfrac{i \omega 4 Z_{12}}{\pi m_2^*}   &  -\omega^2 - \dfrac{4\pi \gamma_2}{U_2^*} i \omega + \left( \dfrac{2\pi}{U_2^*} \right) ^2 - \dfrac{i \omega 4 Z_{22}}{\pi m_2^*}   
    \end{bmatrix},
\end{equation}
the equations \ref{eq:imp_harmonic} can be condensed as the following system
\begin{equation} \label{eq:imp_system}
Z_T \cdot \begin{pmatrix} y_1 \\ y_2  \end{pmatrix} = 0,
\end{equation}
Finally, we define a generalised impedance function as
\begin{equation} \label{eq:imp_function}
H(\omega) = \text{det}(Z_T).
\end{equation}
It is an analytical function of the complex frequency $\omega=\omega_r+i\omega_i$. 

At this point, we can remark that complex roots of \ref{eq:imp_function}, corresponding to nontrivial solutions of the two-dimensional system \ref{eq:imp_system},
also correspond to solutions of the fluid-structure eigenvalue problem  \ref{eq:matricial_eq}. We thus have replaced the resolution of a matricial eigenvalue problem of large dimension by the sole inspection of a $2\times 2$ matrix. This is however a non-linear eigenvalue problem since $\omega$ appears quadratically in $Z_T$ and most importantly because $Z_{ij}$ depends on $\omega$. In practice, the computation of all physically relevant eigenvalues requires the knowledge of the 
functions $Z_{ij}$ in the whole complex $\omega$-plane. However, if one is only interested in localizing the marginally stable states, it is only required to know the values of these functions along the real $\omega$ axis. This property is at the origin of a very efficient method which will be explained and validated in section \ref{sec:valid}, and subsequently used to perform parametric studies in section \ref{sec:param}.


The impedance-based method developed in the present work is an exact realisation of the classical $p-k$ method \citep{hassig1971approximate}, as it provides a fully coupled description of the linear fluid–structure interaction. Instead of relying on approximate aerodynamic transfer functions, the fluid response is obtained directly by solving a series of forced problems, in which the bodies are imposed to oscillate harmonically at prescribed real frequencies. This procedure yields frequency-dependent impedance functions that rigorously encapsulate the hydrodynamic feedback of the flow on the moving structures. The resulting impedance matrix, combined with the structural parameters, defines a compact stability criterion from which the onset of instability can be predicted without explicitly solving the coupled eigenvalue problem. The approach thus retains the interpretability of classical generalised aerodynamic forcing (GAF) based methods while extending the concept to the exact hydrodynamic coupling derived from the full linearised Navier–Stokes equations, enabling accurate and efficient parametric stability analyses across a wide range of configurations.

 The advantage of the criterion described here is that a limited number of calculations is required in order to get the stability prediction of a vast number of different cases. Once a set of forced problems are calculated for a fixed set of \{$Re$ and $L$\}, the stability of systems for any set of  \{$U^*_1$, $U^*_2$, $m^*_1$, $m^*_2$, $\gamma^*_1$, $\gamma^*_2$\} is acquired by the simple inspection of the determinant of $2\times 2$ matrices.

\subsection{Numerical implementation}

The equations are rewritten in a variational formulation, spatially discretised and solved with the FreeFem++ open-source software \citep{hecht2012new}. The problem is necessarily formulated along a truncated domain of sufficient dimension (see figure \ref{fig:CylinderSketch}), and thus the equations are complemented with suitable boundary conditions for the external boundaries ($\Gamma_{in}$, $\Gamma_{l}$, $\Gamma_{out}$). For the fluid variables, a Dirichlet boundary condition is imposed at the inflow boundary $\Gamma_{in}$: $\vec{u}_0 |_{\Gamma_{in}} = U_\infty {\bf e}_x$ with $  U_\infty \equiv 1$, and a stress-free condition is imposed on the lateral and outflow boundaries. For the ALE variables, homogeneous Dirichlet boundary conditions are imposed on all outer boundaries. Following the classical procedure, the base-flow is computed using a Newton method, and the eigenproblems are solved using a shift-invert method as implemented in the SLEPc library.
 
Two additional tricks are employed to lighten the resolution and/or improve the precision. First, mesh adaptation is intensively used to increase the mesh density in regions of strong gradients while decreasing it in other regions.
Secondly, for the linearised problem, we employ the complex mapping method \citep{sierra2020efficient} to characterise the stability properties of the problem and to suppress artificial unstable modes arising due to the strongly convective nature of the wake.  Such a method has been successfully employed in the past in other fluid configurations, for instance, the jet flow past a circular aperture \citep{sierra2022acoustic}, a flow of two coaxial jets \citep{corrochano2023mode} or the wake flow past a rotating particle \citep{sierra2022unveiling}. When using the complex mapping method, the spatial structure of the global mode near the boundary becomes evanescent and does not have an influence on the stability properties of the problem. 

Monitoring of all computations and post-processing is done thanks to the StabFem interface \citep[]{fabre2018practical}. Following to the philosophy of this project, sample codes reproducing key results of the present paper are available on the website of the project \footnote{ \url{https://stabfem.gitlab.io/StabFem/} }.

Regarding the method for threshold detection in terms of the impedance concept explained in 
Sec. \ref{sec:imp}, the numerical resolution procedure consists, in a first step, in generating a tabulation of the impedance functions $Z_{ij}$ as function of $\omega$ and $Re$. Then, zeros of the generalised impedance \ref{eq:imp_function} are computed by considering it as two functions (real and imaginary parts of $H(\omega)$ as function of the two variables $\omega$, $U^*$, and a Levenberg-Marquardt method is used to solve it, using interpolation along the range of tabulated $\omega$ to evaluate the impedances $Z_{ij}$ and their $\omega$-derivatives.

If one looks at the computational cost of traditional Linear Stability Analysis (Arnoldi method) on a standard desktop computer on a single core, solving for $20$ eigenvalues in our configuration takes approximatively $20$ seconds. Let's assume that one needs a total of $10$ computations in order to find the thresholds, which brings a total of $200$ seconds to find the thresholds for a selected set of parameters \{$Re$, $L$, $U^*_1$, $U^*_2$, $m^*_1$, $m^*_2$, $\gamma^*_1$ and $\gamma^*_2$ \}. 
The impedance-based method on the other hand, requires the computation of a set of forced problems, one of them taking approximately $5$ seconds. Considering that one needs tabulated values of $\omega=[0.4:0.01:1.2]$, that brings the computational cost to $405$ seconds. These calculations however, are only needed for a set of \{$Re$ and $L$\}. From there, finding the zeros of the impedance fonction for any set of \{$U^*_1$, $U^*_2$, $m^*_1$, $m^*_2$, $\gamma^*_1$ and $\gamma^*_2$\} is instantaneous.

\section{Validation}
\label{sec:valid}

Throughout this article, we will consider a tandem of spring-mounted cylinders (with the exception of appendix \ref{sec:multiple_bodies}). The Reynolds numbers investigated range up to $Re=100$ and the damping parameters of the cylinders are considered to be the same and noted $\gamma=\gamma_1=\gamma_2$. The damping ratios are set to $0$, except for section \ref{subsec:damp}. The reduced velocity and reduced mass of both cylinders will be considered to be the same and will be noted as $U^*=U_1^*=U_2^*$ and $m^*=m_1^*=m_2^*$, except for part of section \ref{subsec:decup}. 

\subsection{Linear coupled problem}

%
%
\begin{figure}
\centering
\includegraphics[scale=1, trim = 0cm 0cm 0cm 0cm, clip]{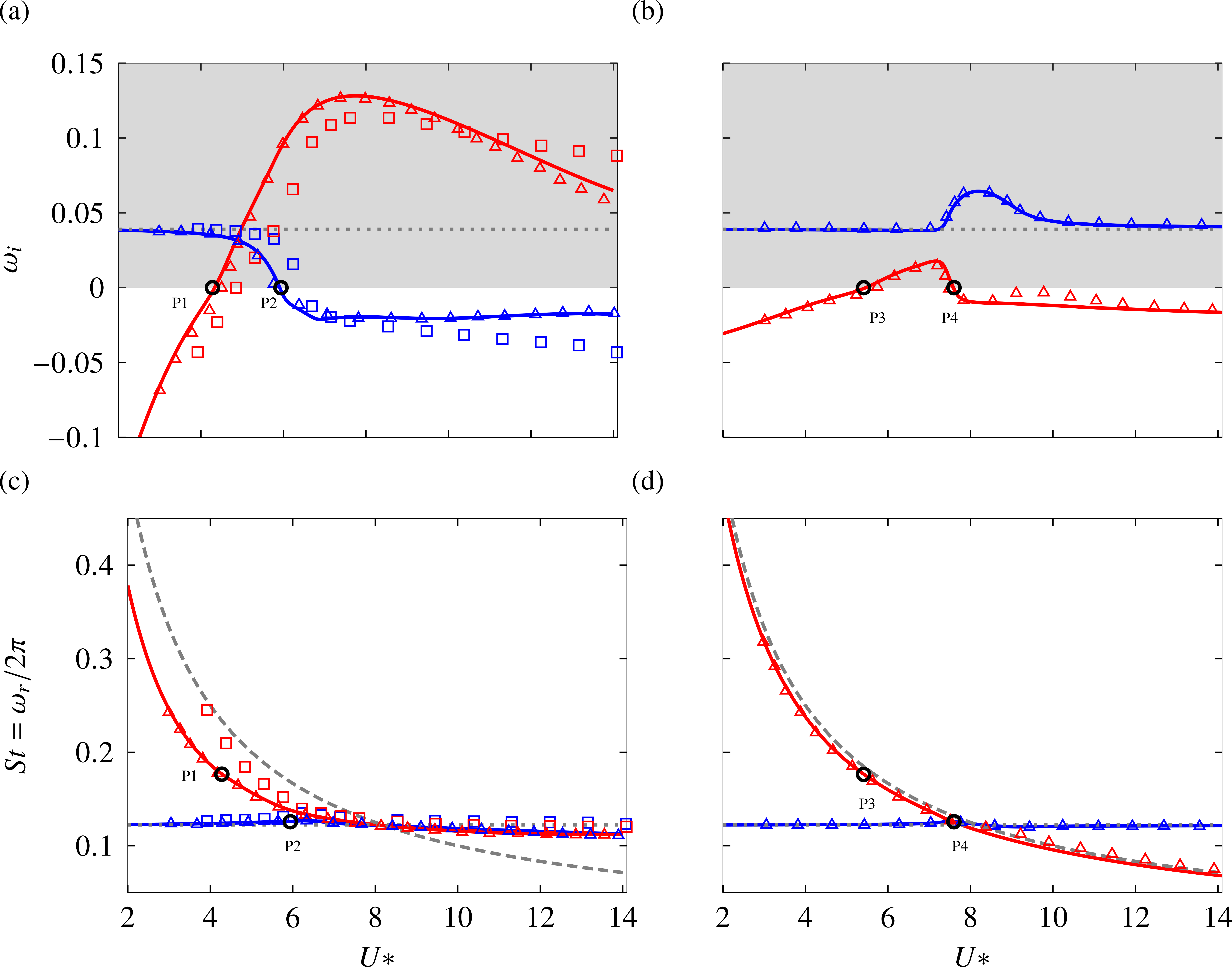}
\caption{Real and imaginary parts of the leading eigenvalues (LSA) with respect to $U^*$ at $Re = 100$ and $L=1.5$ for $m^*  = 2.546$ (a,c) and $m^* = 20$ (b,d). Plain lines are the results of the current study. Results from \citet{tirri2023linear} are shown by \squarered{} and \squareblue{}. Results from \citet{zhang2024global} are shown by \mytriangle{red} and \mytriangle{blue}. The unstable region is depicted as the grey zone. The natural frequency of a spring-mounted cylinder in vacuum $\omega_n=\frac{2\pi}{U_n^*}$ is shown as $\dashedgray{}$. The growth rate and frequency of the fluid mode behind two fixed cylinders are displayed as $\dottedgray{}$. The predictions from the impedance criterion are shown as \hphantom{h}\blackcircle{}.} 
\label{fig:N2_L1.5_Re100}
\end{figure}

Figure \ref{fig:N2_L1.5_Re100} shows the real and imaginary parts of the leading eigenvalues against $U^*$ for  $L=1.5$ at a Reynolds number of  $Re=100$ and for $m^*=2.546$ and $m^*=20$. Two leading unstable modes are found and the evolution of their growth rate and frequency is in very good agreement with \citet{zhang2024global}. For $m^*=2.546$, we observe a slight discrepancy with results from \citet{tirri2023linear}. These authors used an immersed boundary method which was shown to yield incorrect results when the added-mass effect is not properly taken into account \citep[][]{suzuki2011effect}.

\begin{figure}
\centering
\includegraphics[scale=1, trim = 0cm 0cm 0cm 0cm, clip]{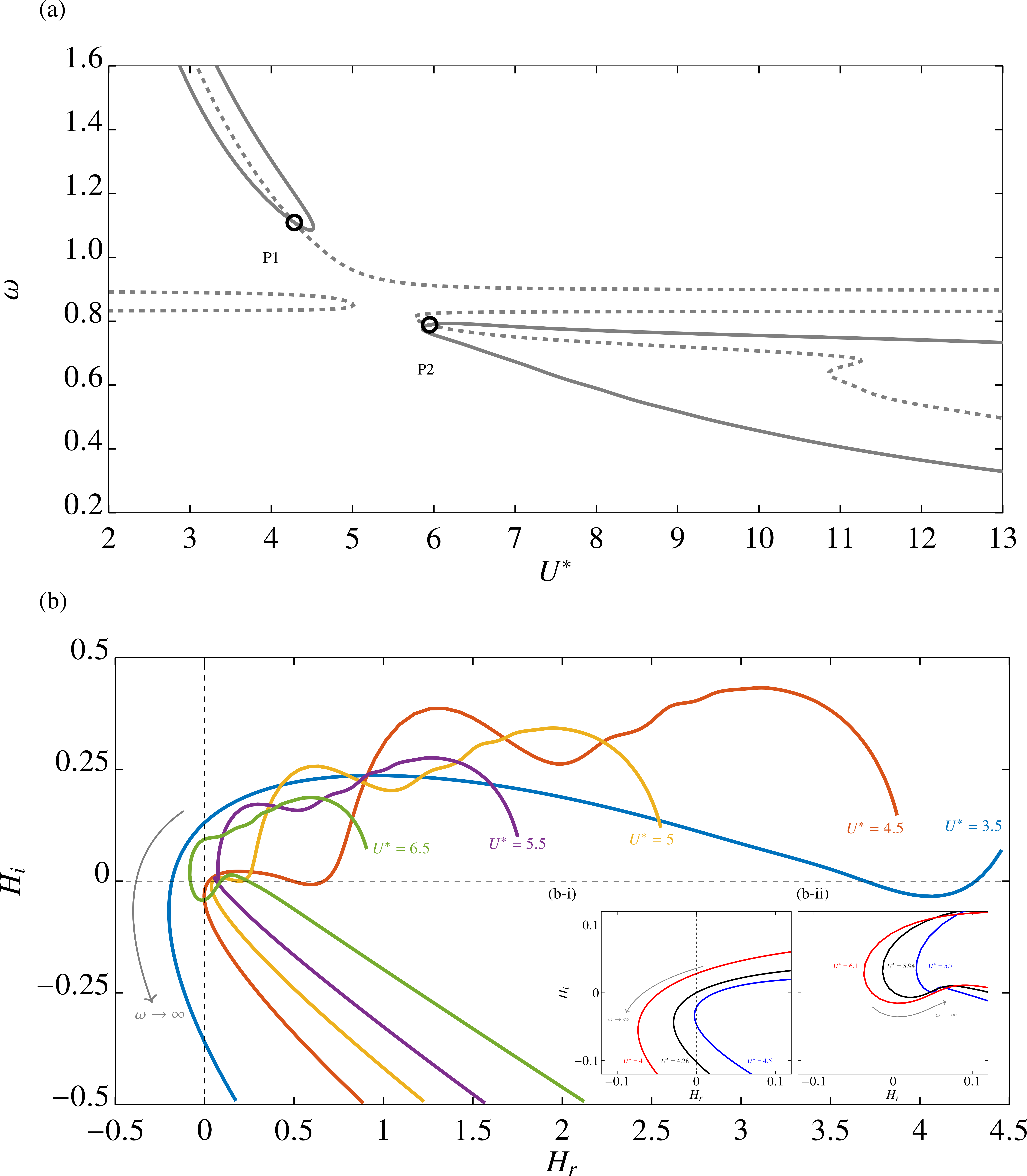}
\caption{Illustration of the threshold detection based on the impedance criterion. (a) Zero isolines of the real ($\fullgray$) and imaginary ($\dashedgray$) parts of the impedance function $H$ in the $\omega - U^*$ plane. The symbols \hphantom{h}\blackcircle{}\hphantom{h} show the zeros of $H$. (b) Plot of the imaginary part of the impedance function $H$ with respect to its real part (Nyquist curve) for several values of $U^*$. Nyquist curves around the points of neutral stability at $U^*=4.28$ (b-i) and $U^*=5.94$ (b-ii).} 
\label{fig:nyquist}
\end{figure}

\subsection{Impedance-based stability predictions}
\label{subsec:valid_imp}

The generalised impedance provides a criterion of instability: the system is unstable if there exists a non-trivial solution of \cref{eq:imp_system}, for which $\omega_i>0$,
as described in \citet{fabre2020acoustic}. The link between impedance and instability can be formulated using Nyquist diagrams. \Cref{fig:nyquist} shows an example of the threshold detection based on impedance criterion for $L=1.5$, $Re=100$ and $m^*=2.546$. \Cref{fig:nyquist}\text{(a)} shows the zero isolines of both the real and imaginary parts of the function $H$ in the $\omega - U^*$ plane. The intersection of these lines gives the predictions of $U^\ast$ and $\omega$ at which the system is neutrally stable (noted as  \blackcircle{}). These predictions are reported in figure \ref{fig:N2_L1.5_Re100}\text{(a)}  and \ref{fig:N2_L1.5_Re100}\text{(c)}  as points P1 and P2, and are in perfect agreement with the results from linear stability. \Cref{fig:nyquist}\text{(b)} shows the Nyquist diagrams, i.e., the imaginary part of the function $H$ against its real part for different values of  $U^*$. For the first point of neutral stability (at $U^*=4.38$), one eigenvalue switches from stable to unstable when increasing the reduced velocity $U^*$, as the Nyquist curve transitions from encircling the origin to having the origin lying at the right of the curve's trajectory, relative to its parametrization from $\omega = 0$ to $\omega \to \infty$ (see figure \ref{fig:nyquist}\text{(b-i)}). However, for the second point of neutral stability (at $U^*=5.94$), one eigenvalue goes from unstable to stable when increasing the reduced velocity $U^*$, and the Nyquist curve transitions from having the origin lying to the right of the curve's trajectory to the curve encircling the origin (see figure \ref{fig:nyquist}\text{(b-ii)}). The impedance-based detection for $m^*=20$ is also in perfect agreement with LSA results as shown in figure \ref{fig:N2_L1.5_Re100}\text{(b)}  and \ref{fig:N2_L1.5_Re100}\text{(d)}  by the points P3 and P4.


\section{Results and discussion}
\label{sec:param}

\subsection{Description of the modes}
\label{subsec:mode_descript}

\begin{figure}
\centering
\includegraphics[scale=1, trim = 0cm 0cm 0cm 0cm, clip]{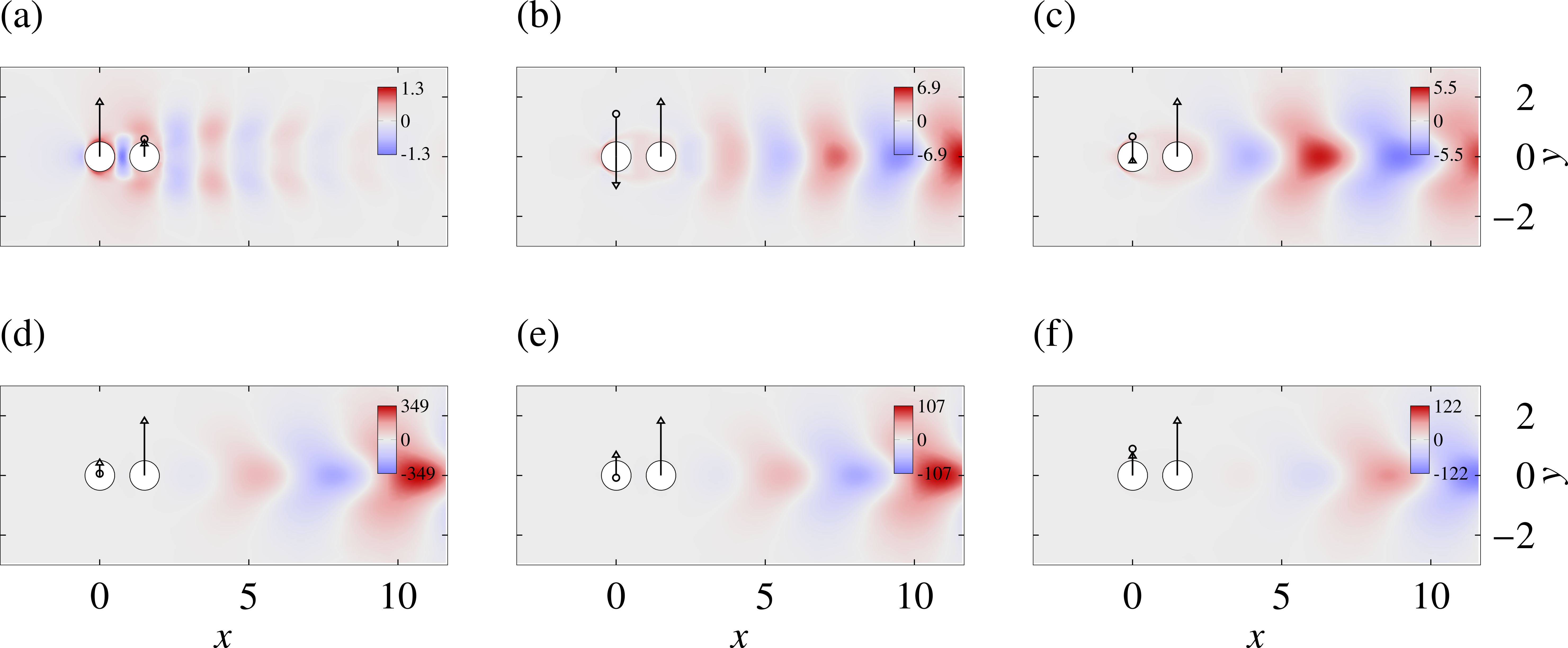}
\caption{Transversal velocity $u_y$ fields of the flow past a tandem of cylinder spaced of $L=1.5$ for $Re=100$ and $m^*=2.546$: mode B (a,b,c) and A (d,e,f) at reduced velocities $U^*=3$ (a,d), $U^*=5$ (b,e) and $U^*=8$ (d,f). See explanations in text for significance of the colour maps and representations of the cylinder's displacement.}
\label{fig:N2_Re100_m2.546_vort}
\end{figure}

Let us first come back to the cases $\{L=1.5; Re=100; m^*=2.546\}$ and $\{L=1.5; Re=100; m^*=20\}$ which were previously used for validation and plotted in figure \ref{fig:N2_L1.5_Re100}, and comment on them from a physical point of view. 
To explain the physical significance of these results, we will also plot figure \ref{fig:N2_Re100_m2.546_vort} the structure of the modes for selected sets of parameters. Each eigenmode has been normalised by setting the highest velocity $z_i$ to $1$. The transversal fluid velocity $u_y$ and the vertical velocities of each cylinder $z_1$ and $z_2$ are shown figure \ref{fig:N2_Re100_m2.546_vort} for $Re=100$ and for $m^*=2.546$. The real and imaginary part of the velocities $z_i$ are respectively plotted as an arrow with a triangle ($\arrowTriangle{}$) and a circle head ($\arrowCircle{}$). For the sake of readability, these amplitudes have been doubled when plotted on the following figures.
With this choice of normalisation, large values of the transversal velocity $u_y$ in the bulk (indicated by the colour ranges of the subplots) imply that the transversal motion of the cylinder associated with the mode is much weaker than the transversal fluid motion. On the other hand, a low maximum in the transversal fluid velocity implies a strong transversal motion of the cylinder.

As reported in \citet{tirri2023linear} and \citet{zhang2024global}, the results of linear stability analysis for $m^*=2.546$ introduces two leading eigenmodes classified as A ($\fullblue{}$) and B ($\fullred{}$). 

The A mode is unstable at low reduced velocities with its growth rate matching that of the leading mode behind the fixed tandem of cylinders $\omega_{f_i}=0.039$ ($\dottedgray{}$ in figure \ref{fig:N2_L1.5_Re100}). It becomes stable at a reduced velocity of $U^*=5.94$ and its frequency varies little and matches the frequency of the fixed tandem mode $\omega_{f_r}=0.76$ ($\dottedgray{}$), with a small decrease towards the highest values of the reduced velocity. This mode has been classified as a "fluidic" mode in the literature. Mode B has a more complex structure arising from the fluid-structure interaction. The mode becomes unstable at $U^*=4.28$ with a maximum growth rate at $U^*=7$. For low reduced velocities, the frequency of the mode matches the natural frequency of a structure-only system ($\omega_n=\frac{2\pi}{U_n^*}$, shown as $\dashedgray{}$ in figure \ref{fig:N2_L1.5_Re100}). For higher reduced velocities, its frequency tends to that of the fixed tandem mode. Both cylinders display transverse motion throughout the whole range of reduced velocities, confirming the structural nature of the mode. In accordance with the results of \citet{tirri2023linear}, the flow field of the A mode resembles that of the wake of two fixed cylinders, as can be seen in figures \ref{fig:N2_Re100_m2.546_vort}\text{(d)}, \ref{fig:N2_Re100_m2.546_vort}\text{(e)} and \ref{fig:N2_Re100_m2.546_vort}\text{(f)}. The large values of the transversal velocity (as seen for the colour ranges used in the subfigure) confirm the fluid-dominated nature of this mode. The region of high transversal velocity is shifted downstream with increasing reduced velocities.
On the other hand, the wake of the B mode includes a high transversal velocity region localised around the bodies, especially at low reduced velocities (as seen in figure \ref{fig:N2_Re100_m2.546_vort}\text{(a)}). At higher reduced velocity, however, the high transversal velocity region is shifted downstream, resembling the structure behind a fixed tandem of cylinders. Correspondingly, for low reduced velocity, the low values of the transversal velocity show that the transversal motion of the cylinder is dominant compared to the fluid's motion (figure \ref{fig:N2_Re100_m2.546_vort}\text{(a)}). Increasing the reduced velocity, the value of the transversal velocity increases slightly as the mode transitions to a more fluid-dominated nature (figures \ref{fig:N2_Re100_m2.546_vort}\text{(b)} and \ref{fig:N2_Re100_m2.546_vort}\text{(c)}). We can also note that generally, the motion of the rear cylinder is greater than of the front, except at low reduced velocity (figure \ref{fig:N2_Re100_m2.546_vort}\text{(a)}). \citet{navrose2016lock} found, for a single oscillating cylinder, that the high frequency of the fluid-elastic modes at low $U^*$ was linked to a high vorticity region close to the body. On the other hand, a low frequency induced a shift of the high vorticity region downstream of the body. We observe the same link between frequency and structure of the wake for the tandem system.

For $m^*=20$ at $Re=100$ (see figures \ref{fig:N2_L1.5_Re100}\text{(b)}  and \ref{fig:N2_L1.5_Re100}\text{(d)} ), the A mode remains unstable for all reduced velocities investigated. Its growth rate matches that of the fixed tandem mode only to depart from it around $7<U^*<10$ with a maximum at $U^*=8.2$. Its frequency is the same as that of the fixed tandem mode $\omega_{f_r}$ and the displacement of both cylinders is null. Concerning B, the mode is unstable for $5.425<U^*<7.6$ with a maximum in growth rate at $7.15$. At $Re=100$, its frequency follows very closely that of the natural structural-only system $\omega_n$, as shown in figure \ref{fig:N2_L1.5_Re100}. The structure of the transversal velocity of the different modes is very similar to that at a lower mass ratio. As the Reynolds number decreases, some exchanges of stability between the modes are observed, as reported in section \ref{subsec:param}.


\subsection{Parametric study}
\label{subsec:param}




\subsubsection{Neutral curves for $m*=2.5$ and $20$; $L = 1.5$ and $3$.}
\label{subsec:NC_valid}

\begin{figure}
\centering
\includegraphics[scale=1, trim = 0cm 0cm 0cm 0cm, clip]{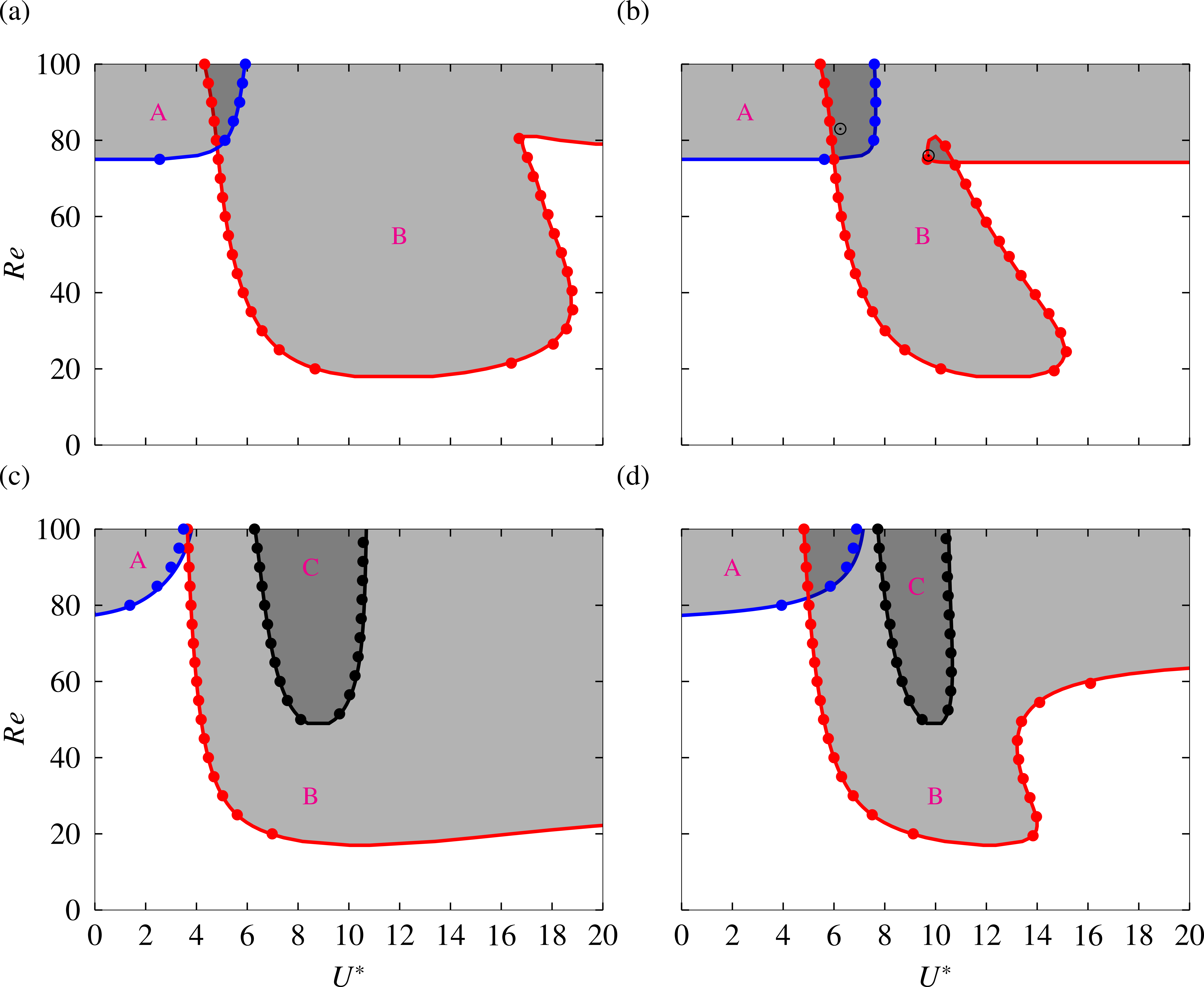}
\caption{ Regions of instability (shaded colors) in the $Re-U^*$ plane for (a,b) $L=1.5$ and (c,d) $L=3$; and for (a,c) $m^*=2.5$ and (b,d) $m^*=20$. Full lines are the results from the impedance-based predictions and markers are results from LSA. Light gray indicates regions where one unstable mode exists and dark gray regions where two unstable modes exist.}
\label{fig:NC_imp_validation}
\end{figure}

Figure \ref{fig:NC_imp_validation} details the neutral stability curves in the $Re-U^*$ plane,  obtained for two values of $m^*$ and two values of $L$. 
Note that the figure displays both results obtained through the resolution of the eigenvalue problem, i.e. LSA, (\myfilledcircle{red}, \myfilledcircle{blue} and \myfilledcircle{black}), and results obtained with the impedance-based method ( $\fullred{}$, $\fullblue{}$ and $\fullblack{}$ ). The excellent agreement of the two methods gives a further validation of the impedance-based method, which will be mostly used in the subsequent section for further parametric studies in a larger range of parameters.
Indeed, as already explained, once the impedance functions have been previously calculated and tabulated, one is able to generate results for all values of the structural parameters with no additional cost than simply solving a $2\times 2$ linear system.

Consider, first, the situation for  $\{L=1.5; m^*=2.5\}$ (figure \ref{fig:NC_imp_validation}\text{(a)}). At this spacing and for a low reduced mass of $m^*=2.5$, the A mode ( \myfilledcircle{blue} ) is stable for all $U^*$ below $Re=75$. Above that Reynolds number, the mode is unstable for values of the reduced velocity below $U^*=6$. 
Concerning B ( \myfilledcircle{red} ), it is unstable for Reynolds numbers above $Re=18$, spanning a broad range of reduced velocities from $U^*\approx4$ to $U^*\approx18$. For Reynolds numbers above $Re=80$, B is unstable for all reduced velocities above $U^*\approx4$. Above $Re\approx78$, the two unstable modes coexist in a region which is indicated by darker gray shading in figure \ref{fig:NC_imp_validation}\text{(a)}.

Consider, now, the case $\{L=1.5,m^*=20\}$ (figure \ref{fig:NC_imp_validation}\text{(b)}). 
For this set of parameters, the mapping of the instability regions is more complex to describe, since an exchange occurs between the two leading branches. Namely, for the largest values of $Re$ considered in the figure, an A mode exists for the whole range of $U^*$ and a B mode exists in a range of $U^*$ centred around the value $U^*=7$. This is consistent with the results previously shown in figure \ref{fig:N2_L1.5_Re100}\text{(b)}  for $Re= 100$ where the threshold of the A branch is displayed in blue while the one of the B branch is displayed in red. When decreasing the Reynolds number, a topological transition occurs, leading to a situation where the A branch for small $U^*$ becomes connected with what was previously the B branch and vice-versa, leading to a situation similar to what was displayed in figure \ref{fig:N2_L1.5_Re100}\text{(a)} . This transition occurs for $Re \approx 84$, at which value one finds a double root of the eigenvalue problem. This exceptional point occurs at $U^* \approx 6.8 $ and is identified by a marker in the figure \ref{fig:NC_imp_validation}\text{(b)}. It is codimension-2 but does not correspond to a bifurcation point, since the coincidence occurs in the unstable region. Due to this exchange of branches, it becomes difficult to distinguish the two modes in the whole range of parameters. 
When decreasing further the Reynolds number, an other peculiar feature appears in the stability map. Namely, the right part of the neutral curve displays a small loop, between the values $Re = 80$ and $Re= 75$. Within this loop, two unstable modes exist. A detailed inspection shows that in this region 
there exists a codimension-2 point where a topological transition occurs. Both exceptional points are represented as $\scriptsize\bullseye$ in figure \ref{fig:NC_imp_validation}\text{(b)}. This complex behaviour is detailed in a supplemental material (see appendix \ref{sec:supp}).

\begin{figure}
\centering
\includegraphics[scale=1, trim = 0cm 0cm 0cm 0cm, clip]{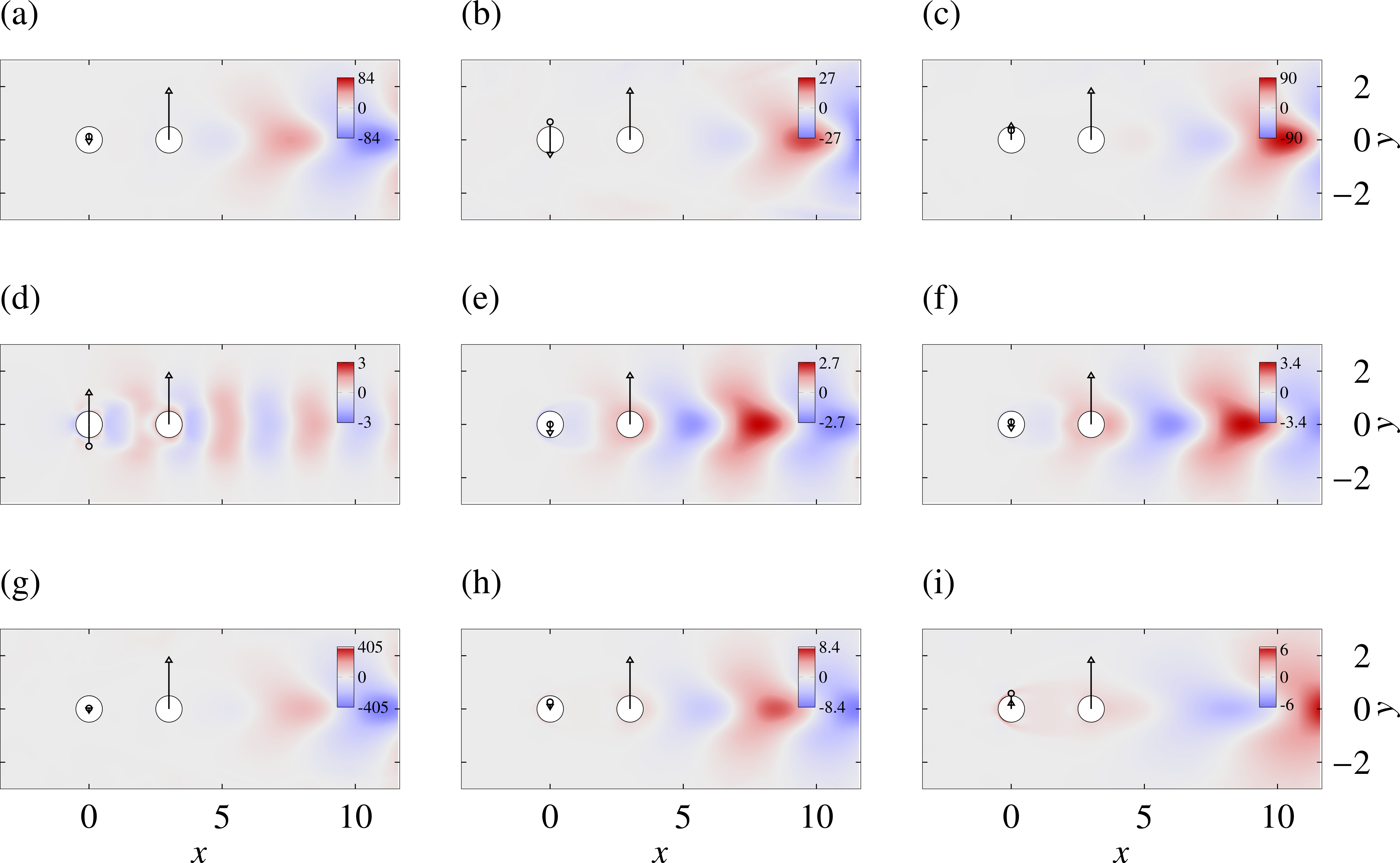}
\caption{Transversal velocity $u_y$ fields of the flow past a tandem of cylinder spaced of $L=3$ for $Re=100$ and $m^*=2.5$: mode A (a,b,c), mode B (d,e,f) and C (g,h,i) at reduced velocities $U^*=4$ (a,d,g), $U^*=8$ (b,e,h) and $U^*=12$ (d,f,i). See explanations in text for the significance of the colour maps and representations of the cylinder's displacement.}
\label{fig:N2_L3_Re100_m2.5_vort}
\end{figure}

Similarly, neutral cuves for $\{L=3;m^*=2.5\}$ and $\{L=3;m^*=20\}$ are respectively plotted in figures \ref{fig:NC_imp_validation}\text{(c)} and \ref{fig:NC_imp_validation}\text{(d)}. 
Three leading eigenmodes are found: A  ( \myfilledcircle{blue} ), B  ( \myfilledcircle{red} ) and C ( \myfilledcircle{black} ). Figure \ref{fig:N2_L3_Re100_m2.5_vort} shows the transversal velocity field of the leading eigenmodes at $m^*=2.5$ and $Re=100$ at different reduced velocities, following the normalisation explained previously. For $m^*=2.5$, A is unstable for $Re>80$ in the low reduced velocity range. At $Re=100$, the frequency of the mode is that of the fixed tandem configuration and the shape of the mode also resembles the wake mode behind the fixed tandem configuration (see figures \ref{fig:N2_L3_Re100_m2.5_vort}\text{(a)} to \ref{fig:N2_L3_Re100_m2.5_vort}\text{(c)}). Moreover, the displacement of both of the cylinders is negligible, as can be seen from the values of the transversal velocity. Mode B, on the other hand, is unstable over a wide range of reduced velocities (above $U^*\approx4$), for $Re>18$. At $Re=100$ and for low reduced velocities, the high transversal velocity region is localised around the front body. Both cylinders exhibit high displacement as seen in figure \ref{fig:N2_L3_Re100_m2.5_vort}\text{(d)}. At the same time, the frequency of the mode follows that of the natural frequency of the structure-only spring-mounted system. Increasing the reduced velocities, the displacement of the rear cylinder becomes greater than the front one and the high transversal velocity region is shifted downstream (see figures \ref{fig:N2_L3_Re100_m2.5_vort}\text{(b)} and \ref{fig:N2_L3_Re100_m2.5_vort}\text{(c)}). The frequency of the mode then tends to that of the fixed tandem configuration. Mode C is unstable for $Re>48$, over a limited range of reduced velocities, between $6<U^*<11$. At $Re=100$, its frequency is that of the fixed tandem configuration.
For low reduced velocities, the shape of the mode also resembles the wake mode behind the fixed tandem configuration (see figure \ref{fig:N2_L3_Re100_m2.5_vort}\text{(g)}). However, towards higher reduced velocities, the rear cylinder exhibits large vertical displacement as can be seen from the low values of the transversal velocity in figures \ref{fig:N2_L3_Re100_m2.5_vort}\text{(h)} and \ref{fig:N2_L3_Re100_m2.5_vort}\text{(i)}. For $m^*=20$, the branch of neutral stability of A is shifted towards higher reduced velocities and the ranges of reduced velocities over which B and C are unstable are reduced.

\subsubsection{Effect of the mass}
\label{subsec:mass}

\begin{figure}
\centering
\includegraphics[scale=1, trim = 0cm 0cm 0cm 0cm, clip]{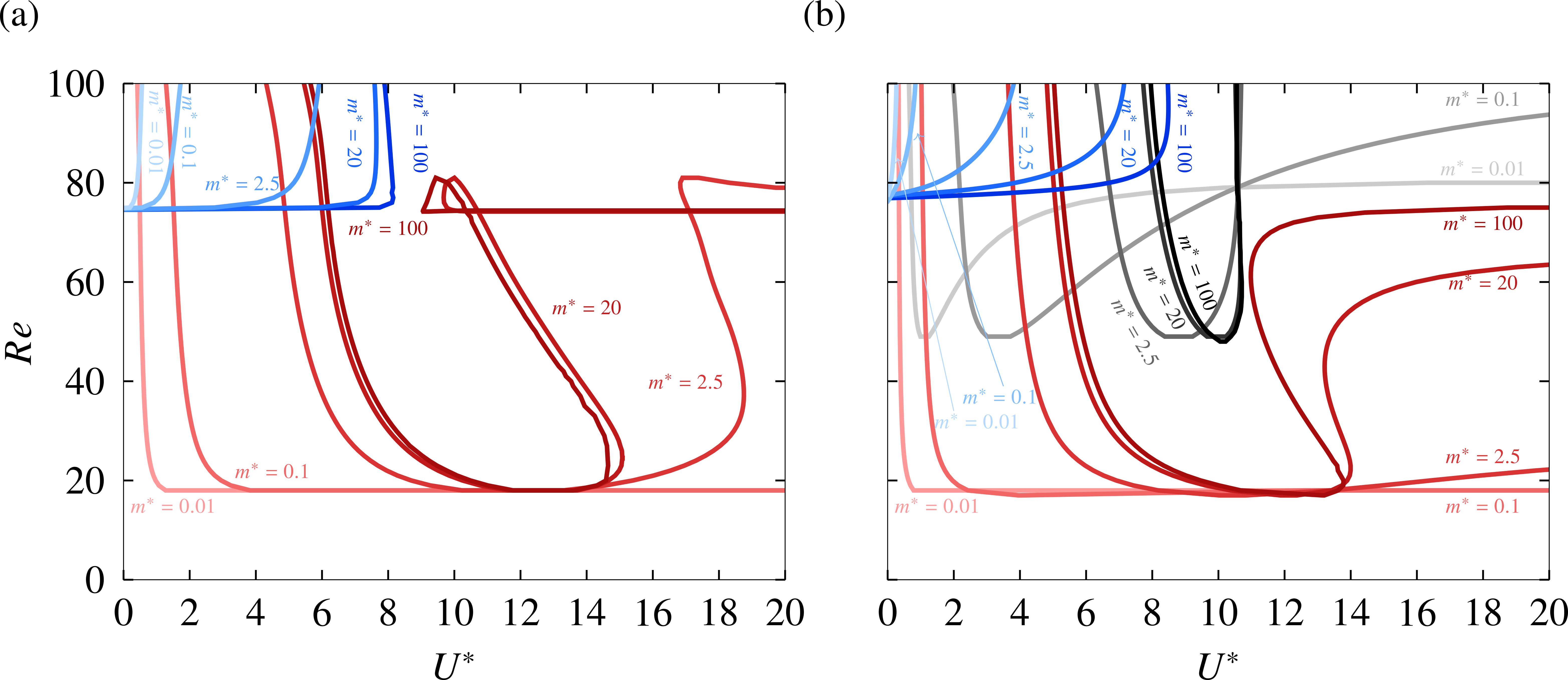}
\caption{ Curves of neutral stability in the $Re-U^*$ plane for (a) $L=1.5$ and (b) $L=3$, computed from the impedance-based method. The colour codes for the different modes are similar to figure \ref{fig:NC_imp_validation}. The colour intensity is going from light to dark for increasing masses whose values are reported in corresponding colours. }
\label{fig:NC_mass}
\end{figure}
After having computed the forced problem for $L=1.5$ and $L=3$ over a range of Reynolds number $Re=[5-100]$ (by steps of $Re=1$), we apply the impedance-based criterion for a range of reduced masses going from $m^*=0.01$ to $m^*=100$. The damping parameter of both cylinders is set to zero. Figure \ref{fig:NC_mass} shows the neutral curves in the $Re-U^*$ plane predicted by the impedance-based criterion. For $L=1.5$ (see figure \ref{fig:NC_mass}\text{(a)}), increasing the reduced mass of the bodies decreases the range of reduced velocities over witch B is unstable. The stability threshold of mode A is shifted towards higher reduced velocities when increasing the reduced mass, extending the range over which the mode is unstable. The exchange of stability observed at $Re \approx 84$ for $m^*=20$ described in section \ref{subsec:NC_valid} is also observed for higher masses.
For $L=3$ (see figure \ref{fig:NC_mass}\text{(b)}), the stability threshold of mode A is also shifted towards higher reduced velocities and the range over which B is unstable is decreased. Concerning mode C, increasing the reduced mass also restricts the range of reduced velocities over which the mode is unstable. Generally speaking, increasing the mass has a stabilising effect on modes B and C and a destabilising effect on mode A.


\subsubsection{Effect of the damping ratio}
\label{subsec:damp}

\begin{figure}
\centering
\includegraphics[scale=1, trim = 0cm 0cm 0cm 0cm, clip]{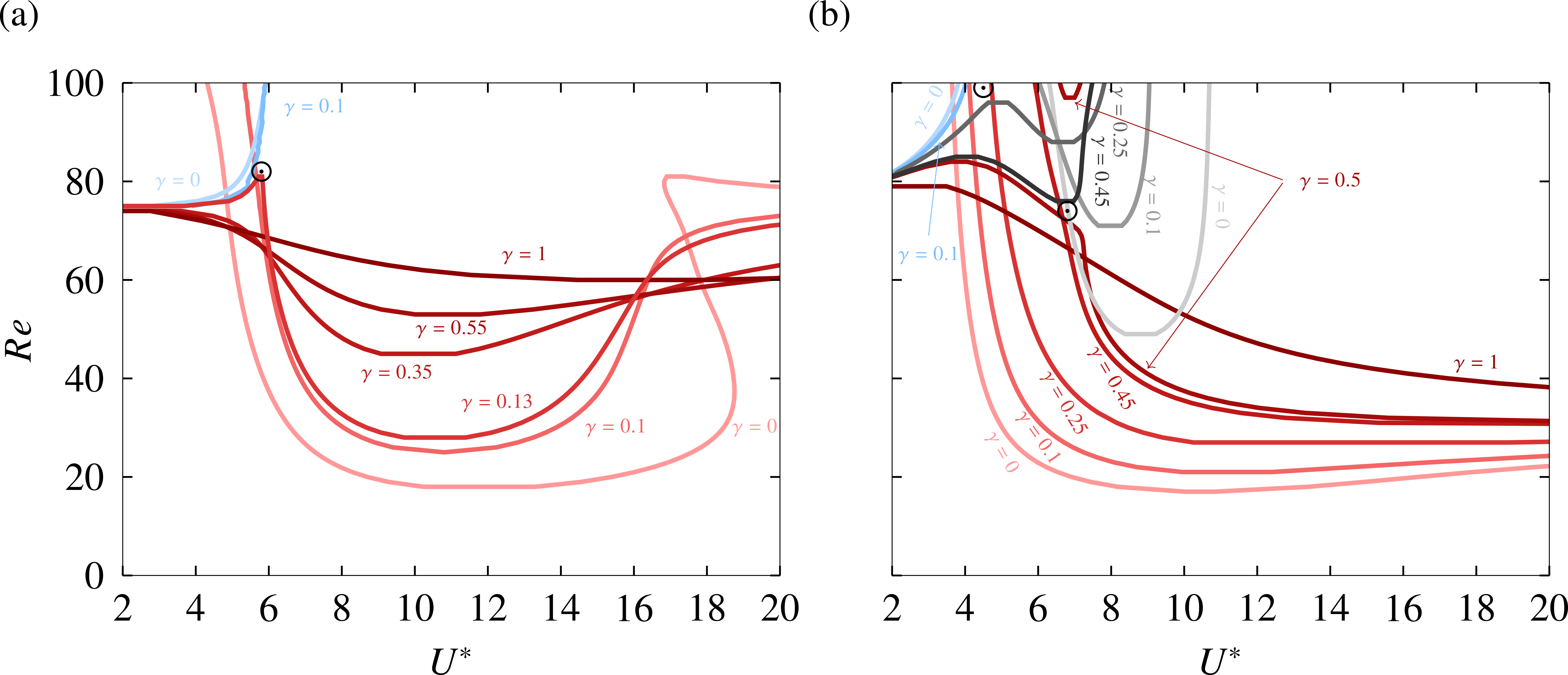}
\caption{ Curves of neutral stability in the $Re-U^*$ plane for a low reduced mass of $m^*=2.5$ and for (a) $L=1.5$ and (b) $L=3$, computed from the impedance-based method. The colour codes for the different modes are similar to figure \ref{fig:NC_imp_validation}. The colour intensity is going from light to dark for increasing damping ratios, whose values are reported in corresponding colours.}
\label{fig:NC_damp}
\end{figure}

\begin{figure}
\includegraphics[scale=1, trim = 0cm 0cm 0cm 0cm, clip]{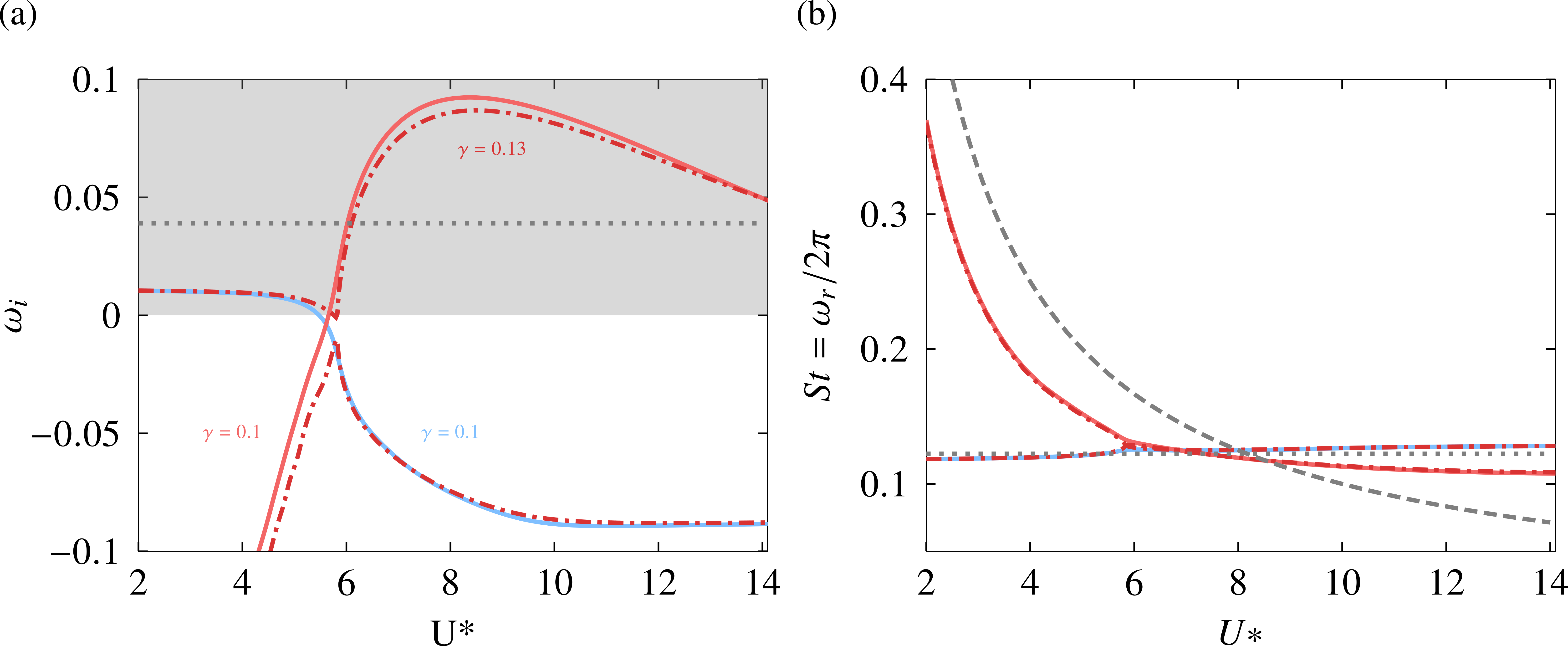}
\caption{Growth rates (c) and frequencies (d) of the leading eigenmodes (LSA) for $L=1.5$ and $m^*=2.5$ at $Re=81$. The modes for $\gamma=0.1$ are plotted with (\fullcustom{red2}) and (\fullcustom{blue2}) and the coalesced mode for $\gamma=0.13$ is plotted with (\fullcustomdashdotted{red3}). The unstable region is depicted as the grey zone. The natural frequency of a spring-mounted cylinder in vacuum $\omega_n=\frac{2\pi}{U_n^*}$ is shown as $\dashedgray{}$. The growth rate and frequency of the fluid mode behind two fixed cylinders are displayed as $\dottedgray{}$.}
\label{fig:NC_damp_codim}
\end{figure}

From the same calculations of the forced problem, the impedance-based criterion is used for a range of damping ratios going from $\gamma=0$ to $\gamma=1$, keeping the reduced mass fixed to $m^*=2.5$. Figure \ref{fig:NC_damp} shows the neutral curves in the $Re-U^*$ plane predicted by the impedance-based criterion. For $L=1.5$ (see figure \ref{fig:NC_damp}\text{(a)}), the A mode remains largely unaffected by light damping ($\gamma=0.1$). On the other hand, the range of reduced velocities over which B is unstable gets narrower with increasing damping. The onset of instability of B is also shifted towards higher Reynolds numbers, going from $Re=18$ for $\gamma=0$ to $Re=60$ for $\gamma=1$. The system exhibits a critical transition at between $\gamma=0.1$ and $\gamma=0.13$ at around $U^*\approx6$ and $Re\approx81$, where the branches coalesce. Figures \ref{fig:NC_damp_codim}\text{(a)} and \ref{fig:NC_damp_codim}\text{(b)} respectively show the growth rates and frequencies of the modes before and after the higher codimension point, for a fixed $Re$ number. The frequency of the coalescing modes being the same at the exceptional point, the latter corresponds to a codimension-3 point (double Hopf with strong resonance) and is plotted as $\scriptsize\bullseye$ in figure \ref{fig:NC_damp}\text{(a)}.

For $L=3$ (figure \ref{fig:NC_damp}\text{(b)}), the threshold of A mode is slightly shifted towards higher reduced velocities by light damping and the ranges over which B and C are unstable are reduced by increasing the damping. Two transitions are observed, which are similar to the one observed for $L=1.5$. Between $\gamma=0.1$ and $\gamma=0.25$, around $Re\approx95$ and $U^*\approx5$, mode A and the low-$U^*$ branch of C coalesce. Similarly, between $\gamma=0.45$ and $\gamma=0.5$, around $Re\approx75$ and $U^*\approx7$, we observe the coalescence of mode B with the low-$U^*$ branch of the previously merged A-C modes. These exceptional points are also of codimension-3 and are plotted as $\scriptsize\bullseye$ in figure \ref{fig:NC_damp}\text{(b)}.

\subsubsection{Effect of the spacing}
\label{subsec:spacing}

\begin{figure}
\centering
\includegraphics[scale=1, trim = 0cm 0cm 0cm 0cm, clip]{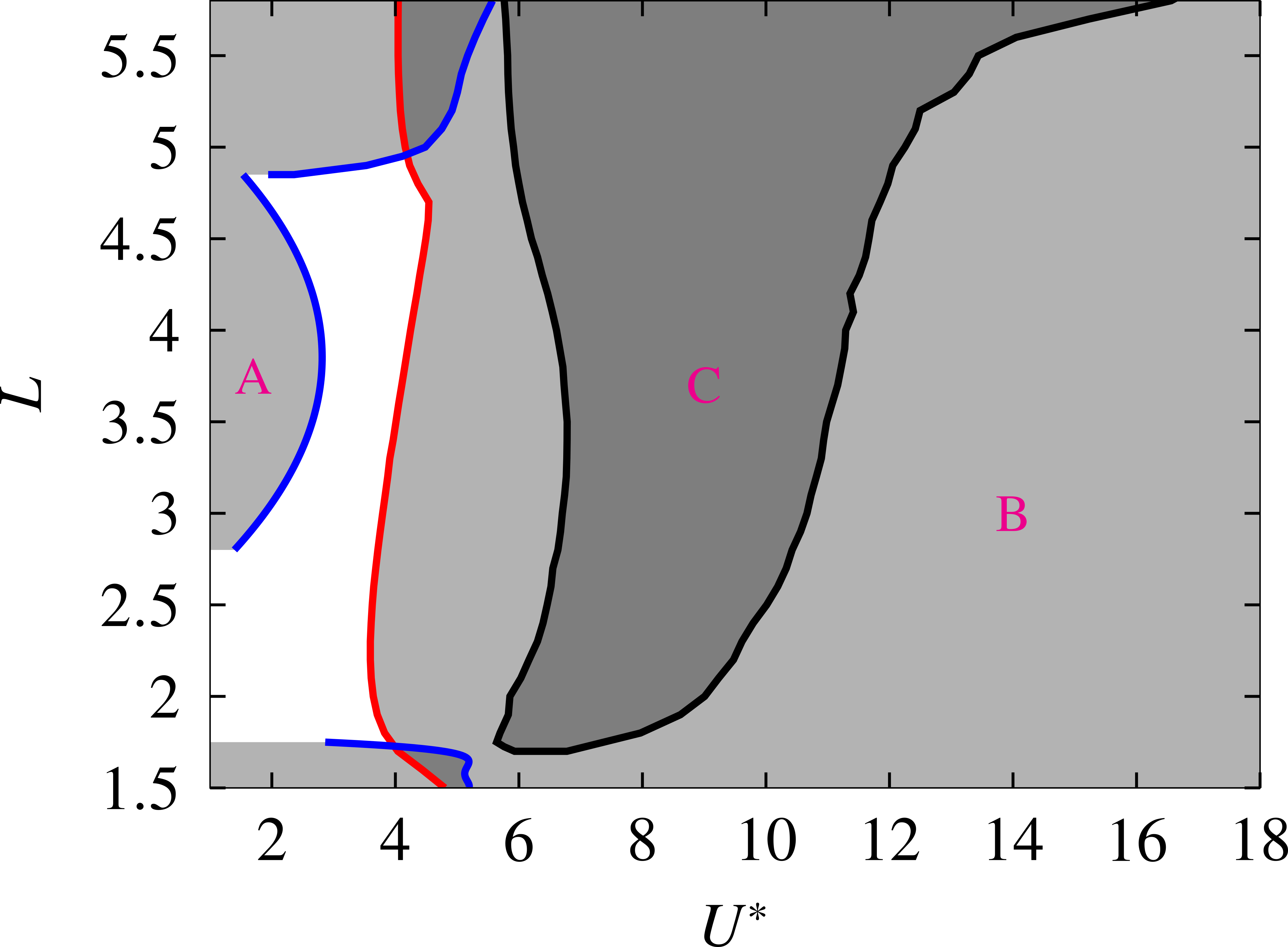}
\caption{Curves of neutral stability in the $L-U^*$ plane for a low reduced mass of $m^*=2.5$ at $Re=80$, computed with the impedance-criterion. The colour codes for the different modes are similar to figure \ref{fig:NC_imp_validation}. Light gray indicates regions where one unstable mode exists and dark gray regions where two unstable modes exist.}
\label{fig:NC_L}
\end{figure}

Figure \ref{fig:NC_L} maps the stability of the tandem of cylinders in the $L-U^*$ plane for $Re = 80$ and $m^*=2.5$. Increasing the distance between the bodies shows dynamics similar to what has been noticed for the fixed-tandem configuration by \cite{zdravkovich1987effects} or by \cite{papaioannou2008effect} for the 2DOF tandem configuration. Mainly, the evolution of mode A, which is foremost a fluid mode, seems to follow different wake interference regimes.
For $1.5<L<1.8$, mode A is unstable in the low $U^*$ range, as described in section \ref{subsec:NC_valid}. This coincides with the "slender body" regime of the fixed tandem, where the free shear layer of the upstream body does not reattach to the downstream one. The vortex shedding comes from the detachment of the upstream body shear layers.
For $1.8<L<3$, mode A is stable and mode C becomes unstable. For the fixed tandem, the shear layers of the front body reattach to the rear one. The vortex street is only formed in the wake of the rear body.
In the $3<L<4.5$ region, A is unstable. This region could correspond to the "intermittent-regime" where the vortex street of the upstream body intermittently reattaches to the rear one.
In this region, the eigenmodes were found to be highly sensitive to variations in the base flow within the gap. Since the computational mesh had to be adapted for each value of the length $L$ in order to solve the forced problem, the resulting neutral curve exhibited some oscillations. To address this, a quadratic fit was applied to smooth the curve.
As of $L>4.5$, the dynamics correspond to the binary vortex street regime where vortex streets are being shed in the wake of both bodies. Up to $L=4.5$, the modes behave in the same way described for $L=3$ and $Re=100$ in section \ref{subsec:valid_imp}. Mainly, mode A induces no displacement of the bodies and modes B and C induce the displacement of both bodies, depending on $U^*$. For spacings above $L=4.5$, however, a change of dynamics is observed. For mode A, both cylinders display motions that are of comparable amplitudes. Mode B induces higher displacement of the front cylinder than of the rear. Finally, C induces higher displacement of the rear cylinder than of the front. Mode B and C start to exhibit a decoupling of the body's motion, whereas mode A arises from the coupling of both cylinders' motion. This phenomenon is confirmed for higher distances, as is explained in the next subsection. 

A possible explanation for the disappearance of mode A in the range $1.8 <L<2.8$ might be that for this range of spacing, the second cylinder lies in the recirculation region of the first one, which corresponds to the region of absolute instability responsible for the instability. A similar explanation was given by \cite{hosseini2021flow} in their investigation of the fixed tandem and three fixed-cylinder configurations for $Re=200$. Accordingly, vortex shedding from the first cylinder is suppressed and the global flow reorganises. In the present case, which involves elastically mounted cylinders, the nature of mode A below $L=4.5$ is dominated by the fluid, so one might expect this scenario to remain essentially valid.

\subsubsection{Mode decoupling at high spacing}
\label{subsec:decup}

\begin{figure}
\centering
\includegraphics[scale=1, trim = 0cm 0cm 0cm 0cm, clip]{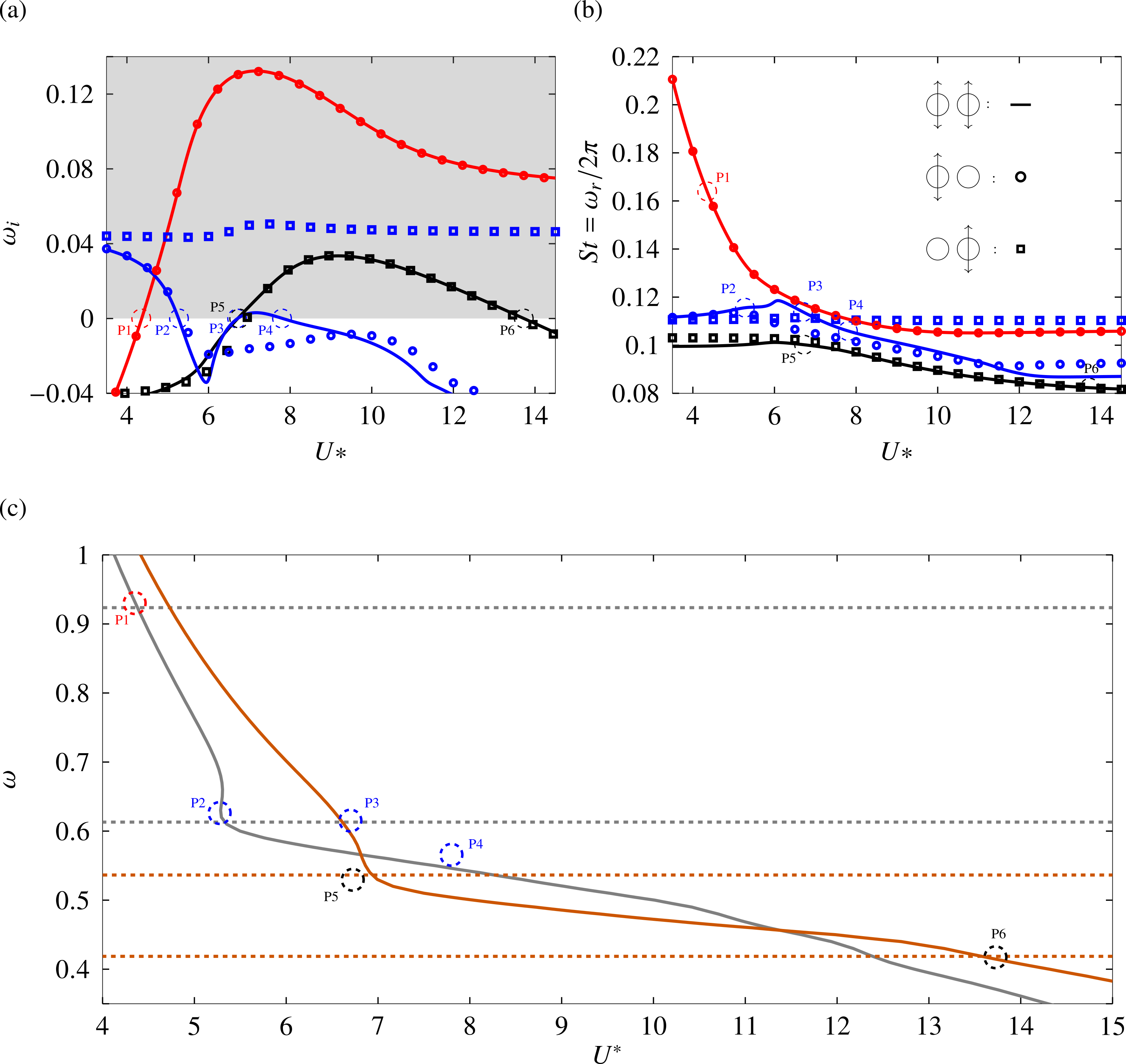}
\caption{Results from LSA of cases with both freely oscillating cylinders ($\fullred{}$, $\fullblue{}$ and $\fullblack{}$), front cylinder fixed (\mysquare{red}, \mysquare{blue} and \mysquare{black}) and rear cylinder fixed (\mycircle{red}, \mycircle{blue} and \mycircle{black}). Real and imaginary parts of the leading eigenvalues with respect to $U^*$ at $Re = 60$ and $L=10$ for $m^* = 2.5$ (a,b). Correspondingly, zero isolines of the real and imaginary parts of $H_1$ (respectively $\fullgray$ and $\dashedgray$) and $H_2$ (respectively $\fullorange$ and $\dashedorange$) in the $\omega - U^*$ plane (c). The points of neutral stability of the different modes for both cylinders freely oscillating are reported in both figures as \mydashedcircle{blue}, \mydashedcircle{red} and \mydashedcircle{black}.}
\label{fig:N2_L10_Re60}
\end{figure}

Figure \ref{fig:N2_L10_Re60} shows results from LSA and from the impedance-based method for $Re = 60$ and $m^* = 2.5$ at $L=10$. Figures \ref{fig:N2_L10_Re60}\text{(a)} and \ref{fig:N2_L10_Re60}\text{(b)} respectively show the growth rate and frequency with respect to $U^*$ for different configurations. Calculations where both bodies are spring-mounted are represented with plain lines. Calculations where either the front or rear cylinder was fixed are respectively shown with circles and rectangles. As can be seen, mode B ($\fullred{}$) of the full problem (both cylinders free to oscillate) is identical to the mode arising from the configuration where the rear cylinder is fixed ( \mycircle{red} ). On the other hand, mode C ($\fullblack{}$) is almost identical to the mode arising from the configuration where the front cylinder is fixed ( \mysquare{black} ). Finally, mode A ($\fullblue{}$) only follows the mode of the rear-fixed cylinder case ( \mycircle{blue} ) for low $U^*$. For $U^*>5.5$, neither mode \mysquare{blue} nor \mycircle{blue} accounts for mode A's dynamics. 

The coupled and decoupled nature of the different modes can also be seen by examination of the terms present in the impedance matrix. Let us define the diagonal terms of $Z_T$ (equation \ref{eq:imp_matrix}) as
\begin{equation}
\begin{cases} 
H_1=  - \omega^2 - \dfrac{4\pi \gamma_1}{U_1^*} i \omega + \left( \dfrac{2\pi}{U_1^*} \right) ^2 - \dfrac{i \omega 4 Z_{1,1}}{\pi m_1^*}, \\[3ex]
H_2 = -\omega^2 - \dfrac{4\pi \gamma_2}{U_2^*} i \omega + \left( \dfrac{2\pi}{U_2^*} \right) ^2 - \dfrac{i \omega 4 Z_{2,2}}{\pi m_2^*}.
\end{cases}
\end{equation}
Figure \ref{fig:N2_L10_Re60}\text{(c)} shows the zero isolines of the real and imaginary parts of $H_1$ (respectively $\fullgray$ and $\dashedgray$) and $H_2$ (respectively $\fullorange$ and $\dashedorange$) in the $\omega - U^*$ plane. The points of neutral stability of the full linear problem (LSA) are reported in figure \ref{fig:N2_L10_Re60}\text{(c)} as dashed circles. One can note that the zeros of $H_1$ approximate relatively well points P1 and P2. As shown from the LSA results, modes B and A are linked to the rear-fixed configuration at these reduced velocities. Similarly, P5 and P6 are relatively well approximated by the zeros of $H_2$. Correspondingly, mode C is linked to the front-fixed configuration at these reduced velocities. On the other hand, the neutral points P3 and P4 are not detected by the sole zeros of $H_1$ or $H_2$, showing there the coupled nature of mode A. Accordingly, the LSA results show that in that range of reduced velocities, A cannot be described from the front-fixed or rear-fixed configurations solely.


\section{Conclusion and perspectives}

The purpose of this paper was to give a thorough description of the linear stability properties of the flow past a tandem of elastically-mounted cylinders, involving both a parametric analysis in a large range of parameters and a physical description of the unstable eigenmodes. 

Due to the large number of parameters involved in the problem, one was led to design an efficient method to perform parametric analyses. For this sake, we designed a generic method which consists of solving in a first step a series of {\em forced problems} in which the cylinders are imposed to harmonically oscillate at a given real frequency $\omega$. This allows us to define {\em impedance functions} $Z_{ij}(\omega)$, condensing in a synthetic way the retroaction of the fluid onto the oscillating bodies. This allows, in a second step, to predict the instability criteria by the sole inspection of a $2\times2$  matrix involving the parameters of the structure (reduced mass, velocity and damping) along with these impedance functions.
  
The impedance-based method was compared to the resolution of the eigenvalue problem for the coupled fluid-body problems, yielding very good agreement. Using the impedance-based stability criterion, an extensive parametric study was then conducted. The effect of increasing the mass and the damping ratio are found to be generally stabilising, except for mode A which is destabilised by the increase of the mass. 
The stability analysis in the $L-U^*$ plane for $ Re = 80$ and $m^* = 2.5$ reveals that the tandem cylinder dynamics evolve with spacing, showing similarities to previously observed wake interference regimes. Mode A dominates in certain regions and aligns with fluid instabilities, while transitions between regimes reflect changes in wake interactions, from slender body behaviour to binary vortex shedding. For $L > 4.5 $, a shift in dynamics occurs, with modes B and C showing decoupled motions of the cylinders, while A remains a coupled mode. This is confirmed by further analysis at $ Re = 60$ and $ L = 10$, where comparison with fixed-body configurations shows that modes B and C align with the rear-fixed and front-fixed cases, respectively. A, however, cannot be captured by either configuration alone at higher reduced velocities, underscoring its coupled nature, as supported by the impedance matrix analysis.

A number of perspectives are opened for the continuation of this study. First, the formalism developed here applies to an arbitrary number of rigid bodies, with almost identical computational cost compared to the fluid problem owing to the use of the discrete-ALE ansatz.  An illustration for a system of three cylinders is given in the appendix \ref{sec:multiple_bodies}. Extension to a higher number of bodies is straightforward. A particularly interesting perspective would be to consider periodic configurations, for which one can expect to apply Floquet-Bloch approaches allowing one to predict the large-scale dynamics of such configurations based on the resolution of problems simply formulated into an elementary cell. 

Aside from increasing the number of bodies, extending to additional degrees of freedom, including streamwise oscillation and rotational motion, could also be considered. According to the discussion in the introduction, such degrees of freedom are not expected to lead to novel dynamics in the case of tandem bodies for symmetry reasons, as in-line oscillations would be coupled to symmetric vortex shedding which is unlikely to occur. On the other hand, streamwise motions could come into play for configurations such as the side-by-side cylinders or the three-cylinder in pinball configuration investigated in appendix \ref{sec:multiple_bodies}. In such a case, streamwise oscillations in an antisymmetric manner could happen, especially in configurations where the cylinders are close enough to lead to a global antisymmetric vortex shedding similar to that of a single composite body \citep{carini2014first}.

Another perspective would be to investigate the dynamics of a set of oscillating cylinders located below a free surface. Recent studies \citep{patel2025coupled} have investigated the effect of a free surface on the vortex-shedding instability behind a fixed cylinder. Interesting dynamics can be anticipated by considering spring-mounted objects. This topic is currently the object of ongoing investigations in our team.

Lastly, since the present study was restricted to the linear regime, one should extend the study by considering nonlinear dynamics. Considering the nonlinear dynamics is especially interesting for two main reasons. First, in the course of the parametric study, a number of exceptional points of codimension 2 or 3 have been identified. Rich dynamics can be expected in the vicinity of such points, which could be elucidated thanks to nonlinear simulations or alternative approaches such as weakly nonlinear expansions, Time Spectral Method or Spectral Submanifolds.
Secondly, considering the application to energy harvesting which was the original motivation for this work, evaluation of the performances and energy efficiencies necessarily requires investigation of the nonlinear regimes. A key question for such applications will be to identify the optimal range of parameters, reduced mass and velocity, and most importantly, the parameter $\gamma_i$ identified here with a damping parameter, but which for an energy harvester would rather be identified as an energy extraction coefficient.



%
%
%


\backsection[Acknowledgements]{Théo Mouyen would like to thank Edouard Boujo for sharing unpublished results and exchanging thoughts during the validation process of the L-ALE method presented in this paper.}


\backsection[Declaration of interests]{The authors report no conflict of interest.}

\backsection[Data availability statement]{The codes used to produce the results in this paper are available in the StabFEM open source project.}

\backsection[Author ORCIDs]{T. Mouyen, https://orcid.org/0000-0001-9590-9126; J. Sierra-Ausin, https://orcid.org/0000-0009-8765-4321; D. Fabre, https://orcid.org/0000-0003-1875-6386; F. Giannetti, https://orcid.org/0000-0002-3744-3978}


\noindent This is an Open Access article written under the terms of the CC-BY 4.0 licence (https://creativecommons.org/licenses/by/4.0/).

\appendix

\section{Exchange of stability between mode branches at $m^*=20$}
\label{sec:supp}

\begin{figure}
\centering
\includegraphics[scale=1, trim = 0cm 0cm 0cm 0cm, clip]{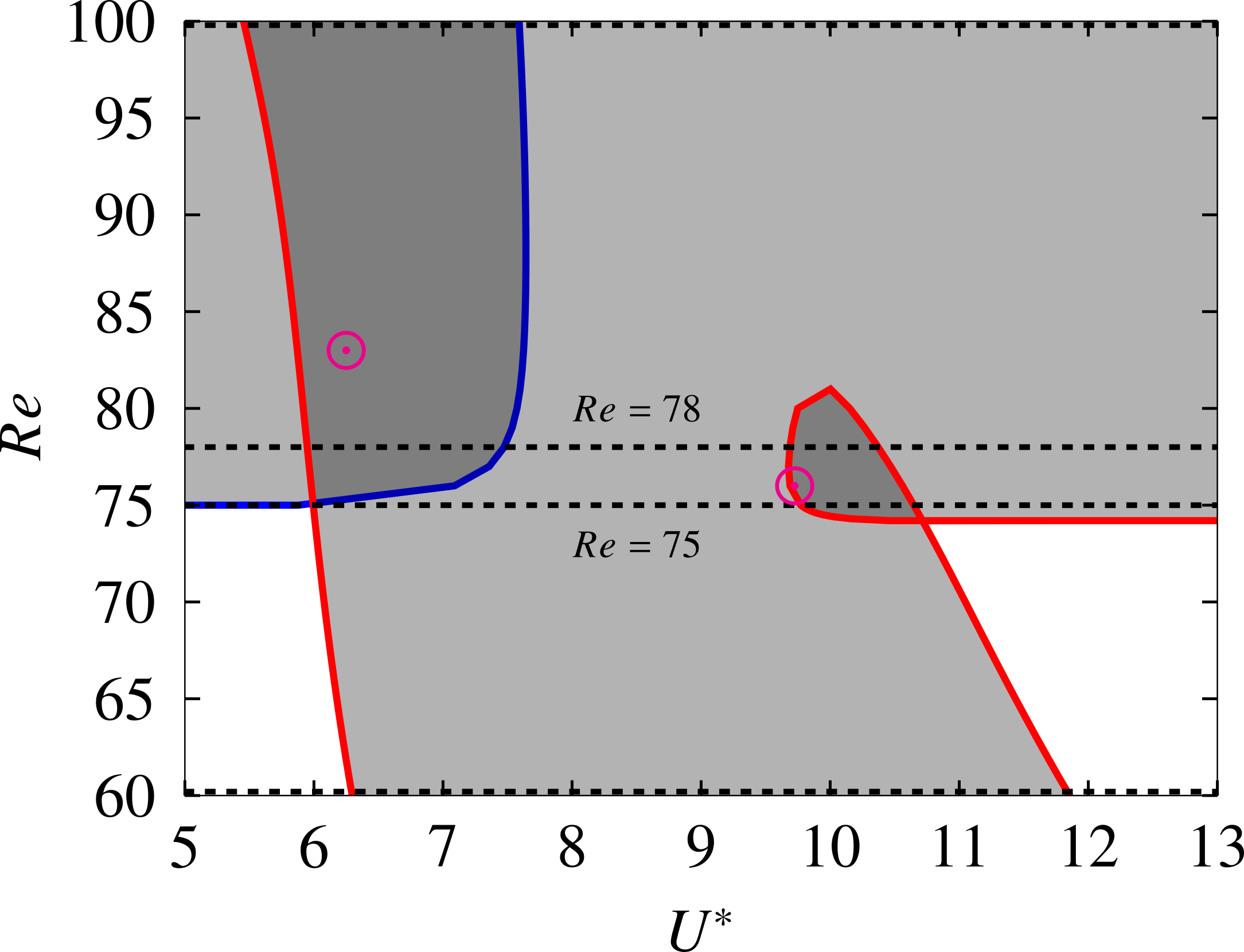}
\caption{Curves of neutral stability in the $Re-U^*$ plane for $m^*=2.5$ and $L=1.5$, from the impedance-based method (zoom of figure \ref{fig:NC_imp_validation}\text{(b)}). The exceptional points are plotted as $\scriptsize\bullseyemagenta$.}
\label{fig:zoom_Nc_m20}
\end{figure}

In section \ref{subsec:param} we showed that for $m^*=20$, multiple exchanges of stability between the modes occur. To clarify the dynamics taking place, we show the growth rates, frequencies as well as root loci from of the different modes from LSA (figure \ref{fig:supp_cont}) for different relevant Reynolds numbers (plotted as $\dashedblack{}$ figure \ref{fig:zoom_Nc_m20}, which is a zoom of figure \ref{fig:NC_imp_validation}\text{(b)}). The points of neutral stability have been plotted firgure \ref{fig:supp_cont} as \hphantom{h}\redcircle{}\hphantom{h} or \hphantom{h}\bluecircle{}\hphantom{h}, according to the colours of the neutral curves in figure \ref{fig:zoom_Nc_m20}.

When decreasing the Reynolds number, the first exchange occurs around $Re\approx84$. The A branch for small $U^*$ becomes connected with what was previously the B branch and vice-versa, as can be seen comparing figure \ref{fig:supp_cont}\text{(a)} for $Re=100$ and figure \ref{fig:supp_cont}\text{(d)} for $Re=78$. This exceptional point occurs for $U^*\approx6.8$ and is identified figure \ref{fig:zoom_Nc_m20} as $\scriptsize\bullseyemagenta$.

The second exchange appears in the loop feature of the neutral curve, around $Re\approx75.5$. For $U^*\approx9.75$ (second symbol $\scriptsize\bullseyemagenta$ in figure \ref{fig:zoom_Nc_m20}), the two unstable branches connect and exchange stability behaviour as can be seen comparing figures \ref{fig:supp_cont}\text{(d)} and \ref{fig:supp_cont}\text{(g)}. The asymptotic limit for low $U^*$ at $Re=75$ corresponds to the critical Reynolds number for which the mode of the fixed tandem configuration is neutral (see $\dottedgray{}$ figure \ref{fig:supp_cont}\text{(g)}).

Note that all neutral points plotted as \hphantom{h}\redcircle{}\hphantom{h} occur when the modes are structure-dominated, as can be seen by looking at the matching frequency. On the other hand, all neutral points plotted as \hphantom{h}\bluecircle{}\hphantom{h} arise from fluid-dominated modes.

\begin{figure}
\centering
\includegraphics[scale=1, trim = 0cm 0cm 0cm 0cm, clip]{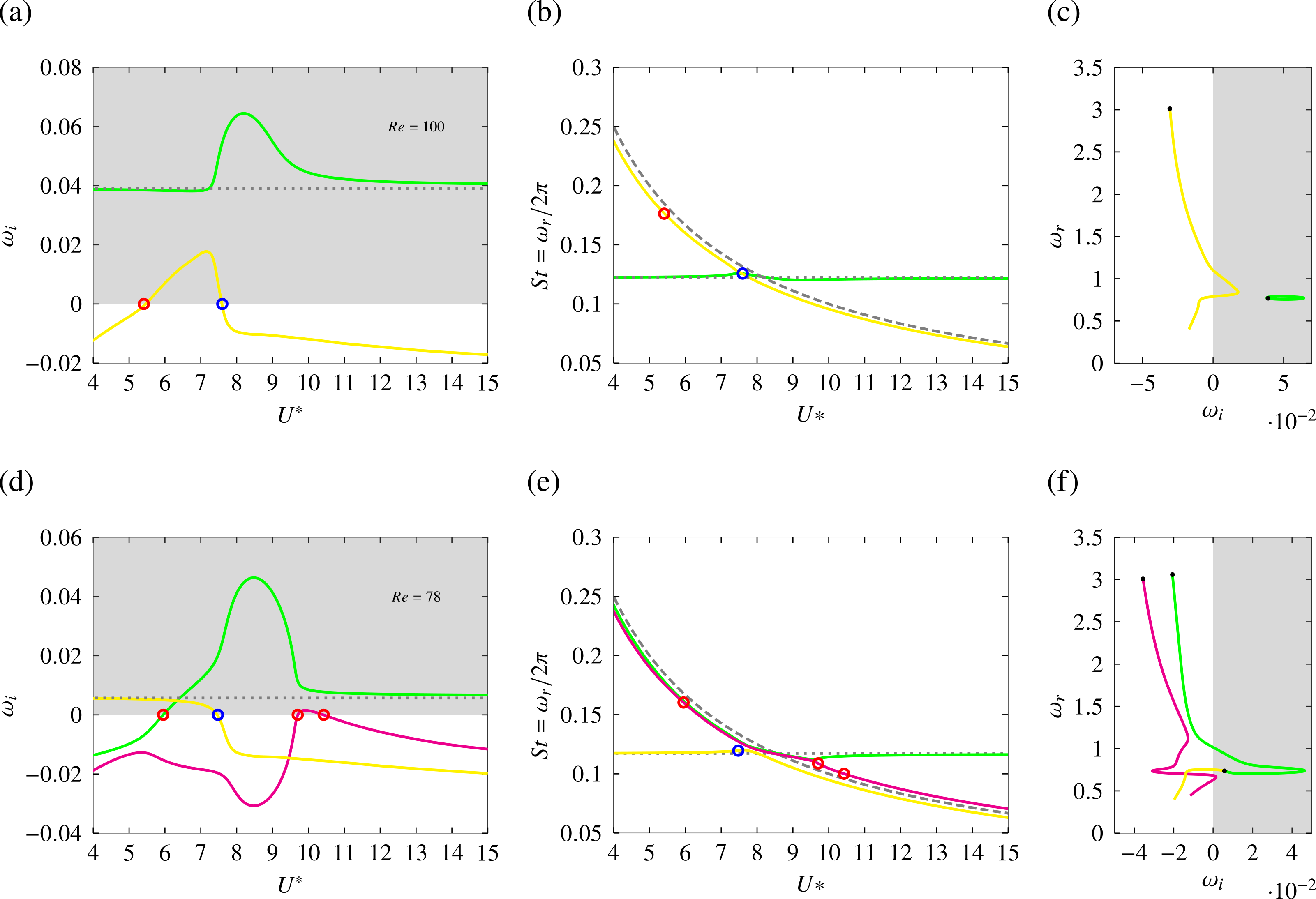}
\caption{ Growth rates (a,d,g,j), frequencies (b,e,h,k) and root loci (c,f,i,l) of the leading eigenmodes (LSA) for $m^*=20$ and $L=1.5$: $Re=100$ (a,b,c), $Re=78$ (d,e,f), $Re=75$ (g,h,i) and $Re=60$ (j,k,l). The natural frequency of a spring-mounted cylinder in vacuum $\omega_n=\frac{2\pi}{U_n^*}$ is shown as $\dashedgray{}$. The growth rate and frequency of the fluid mode in the wake of the corresponding fixed configuration are displayed as $\dottedgray{}$.}
\label{fig:supp_cont}
\end{figure}

\begin{figure}\ContinuedFloat
\centering
\includegraphics[scale=1, trim = 0cm 0cm 0cm 0cm, clip]{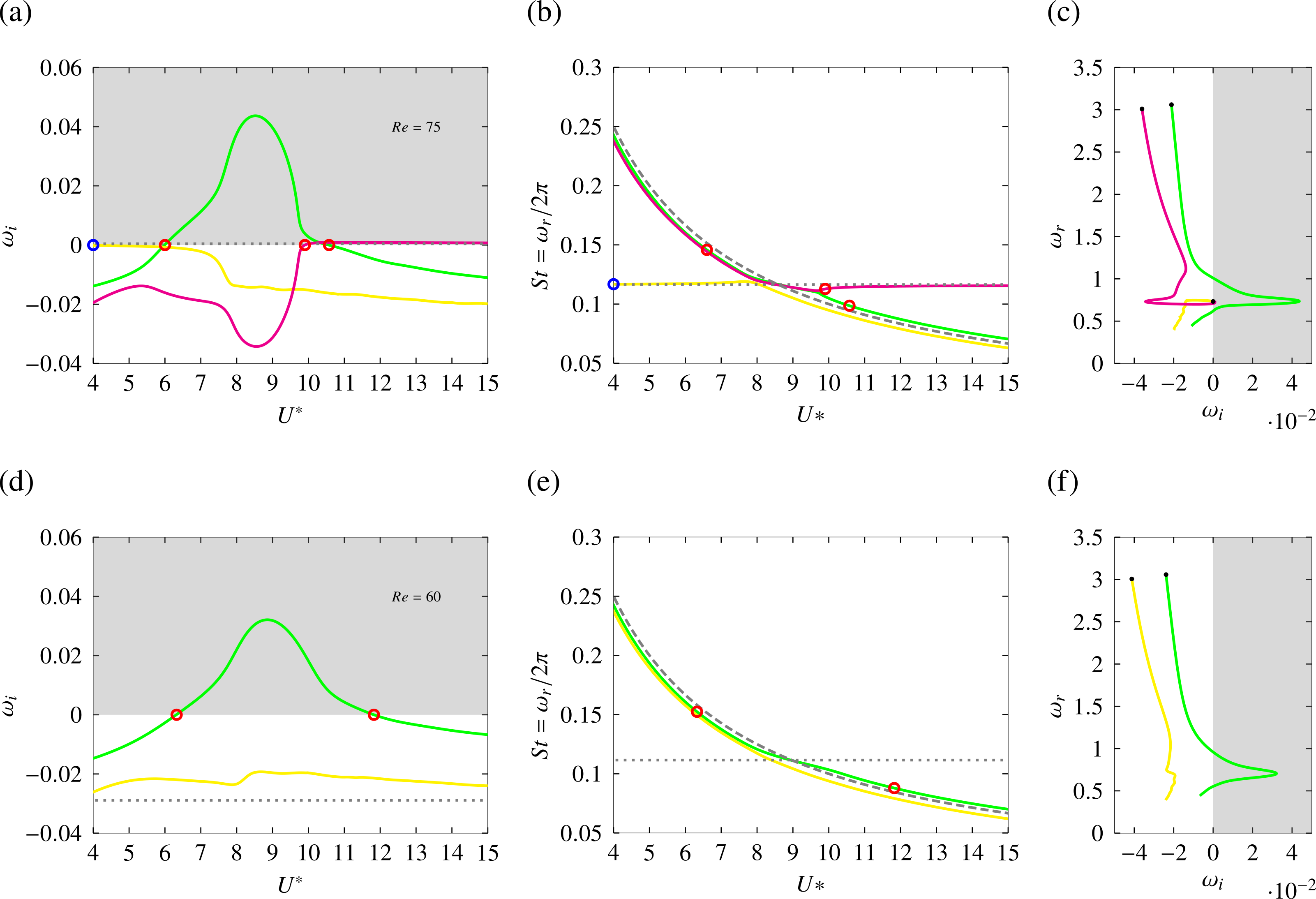}
\caption{ Growth rates (a,d,g,j), frequencies (b,e,h,k) and root loci (c,f,i,l) of the leading eigenmodes (LSA) for $m^*=20$ and $L=1.5$: $Re=100$ (a,b,c), $Re=78$ (d,e,f), $Re=75$ (g,h,i) and $Re=60$ (j,k,l). The natural frequency of a spring-mounted cylinder in vacuum $\omega_n=\frac{2\pi}{U_n^*}$ is shown as $\dashedgray{}$. The growth rate and frequency of the fluid mode in the wake of the corresponding fixed configuration are displayed as $\dottedgray{}$. (cont.)}
\label{}
\end{figure}

\section{Impedance-based criterion of a 3-body system}
\label{sec:multiple_bodies}

\begin{figure}
\centering
\includegraphics[scale=1, trim = 0cm 0cm 0cm 0cm, clip]{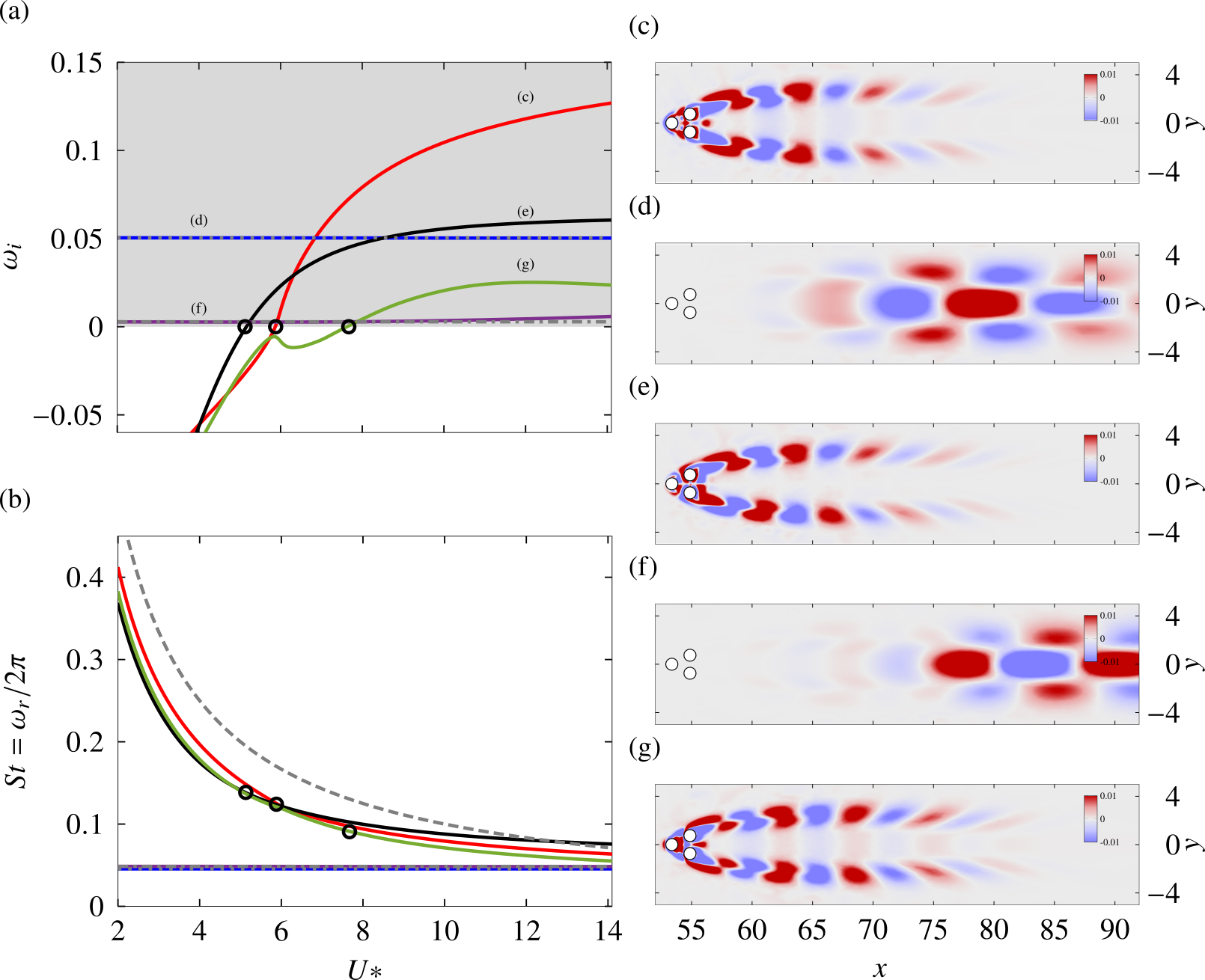}
\caption{Real part of the leading eigenvalues from LSA (a) and their frequencies (b) with respect to $U^*$ at $Re = 60$ for $m_1^* = m_2^* = m_3^* = 2.5$.
The unstable region is depicted as the grey zone. The natural frequency of a spring mounted cylinder in vacuum $\omega_n=\frac{2\pi}{U_n^*}$ is shown as $\dashedgray{}$. The two leading eigenvalues of the corresponding fixed tandem case are displayed as $\dottedgray{}$ and $\dashdottededgray{}$. The predictions from the impedance criterion are shown as \blackcircle{}. Vorticity fields of the eigenmodes (c,d,e,f,g) are showed for $U^*=7$.}
\label{fig:N3}
\end{figure}

The impedance-based stability prediction can be applied to any number $n$ of bodies. We show here the validity of the method for the prediction of the stability of a cluster of three cylinders that are centred on the vertices of an equilateral triangle of side length $3D/2$. The cylinders are free to oscillate in the transverse direction of the flow.  For fixed cylinders, \citet{chen2020numerical} showed that the flow dynamics were highly sensitive to the spacing $L$ and Reynolds number $Re$. This configuration is also known as the fluidic pinball when the cylinders are rotatable. The linear stability analysis at $Re=60$ for $m^*=2.5$ introduces five leading eigenmodes, two being fluid-dominated modes while the three others are of structure-dominated nature. Figure \ref{fig:N3} shows the real and imaginary parts of the leading eigenvalues against $U^*$ (respectively \ref{fig:N3}\text{(a)} and \ref{fig:N3}\text{(b)}) as well as the vorticity field of the different modes at a reduced velocity of $U^*=7$ (figures \ref{fig:N3}\text{(c)},\ref{fig:N3}\text{(d)},\ref{fig:N3}\text{(e)},\ref{fig:N3}\text{(f)} and \ref{fig:N3}\text{(g)}). 
Modes A (\fullblue) and B (\fullpurple) are both unstable over the whole range of reduced velocities investigated and their frequencies respectively match the frequencies of the two unstable modes found in the corresponding fixed case (displayed as $\dottedgray{}$ and $\dashdottededgray{}$). These modes do not induce displacement of any of the three bodies. Their vorticity fields are respectively plotted in figures \ref{fig:N3}\text{(d)} and \ref{fig:N3}\text{(f)} at reduced velocity $U^*=7$. Modes C (\fullred), D (\fullblack) and E (\fullgreen) respectively become unstable at reduced velocities $U^*=5.88$, $U^*=5.13$ and $U^*=7.67$. All three modes follow the natural frequency of the structure-only spring-mounted system ($\omega_n=\frac{2\pi}{U_n^*}$, shown as $\dashedgray{}$). Modes C and E induce the displacement of all cylinders. Both eigenmodes are symmetric and the zone of high vorticity is localised around the bodies as well as in the outer regions of the wake in proximity to the bodies, as seen in figures \ref{fig:N3}\text{(c)} and \ref{fig:N3}\text{(g)}. Mode D, on the other hand, induces the displacement of both rear cylinders, leaving the front one stationary. The structure of the mode is similar to that of the other structure-dominated modes but it is antisymmetric, as can be seen in figure \ref{fig:N3}\text{(e)}.
We computed the forced problem for that configuration and generalised the impedance-based criterion described in section \ref{sec:imp} to a three-body problem. Practically, the thresholds are found by the inspection of the determinant of a $3\times3$ matrix. The stability predictions from the impedance-based criterion are plotted as \blackcircle{} in figures \ref{fig:N3}\text{(a)} and \ref{fig:N3}\text{(b)}. The model correctly predicts the reduced velocities at which the growth rates of the modes become neutral. The prediction of the modes' frequencies is also in very good agreement with the results from LSA.

\bibliographystyle{jfm}

\bibliography{FSI.bib,num_schemes.bib,linear_stab.bib,bluff_bodies.bib,jets.bib,free_surface.bib}

\begin{thebibliography}{49}
\expandafter\ifx\csname natexlab\endcsname\relax\def\natexlab#1{#1}\fi
\def\au#1{#1} \def\ed#1{#1} \def\yr#1{#1}\def\at#1{#1}\def\jt#1{\textit{#1}}
  \def\bt#1{#1}\def\bvol#1{\textbf{#1}} \def\vol#1{#1} \def\pg#1{#1}
  \def\publ#1{#1}\def\arxiv#1{#1}\def\org#1{#1}\def\st#1{\textit{#1}}

\bibitem[Alam(2012)]{alam2012nonlinear}
{\sc \au{Alam, Mohammad-Reza}} \yr{2012}  \at{Nonlinear analysis of an actuated
  seafloor-mounted carpet for a high-performance wave energy extraction}.
  \jt{Proceedings of the Royal Society A: Mathematical, Physical and
  Engineering Sciences}  \bvol{468}~(2146),  \pg{3153--3171}.

\bibitem[Assi {\em et~al.\/}(2013)Assi, Bearman, Carmo, Meneghini, Sherwin \&
  Willden]{assi2013role}
{\sc \au{Assi, GR da~S}, \au{Bearman, Peter~W}, \au{Carmo, Bruno~Souza},
  \au{Meneghini, Julio~Romano}, \au{Sherwin, Spencer~J} \& \au{Willden, RHJ}}
  \yr{2013}  \at{The role of wake stiffness on the wake-induced vibration of
  the downstream cylinder of a tandem pair}.  \jt{Journal of Fluid Mechanics}
  \bvol{718},  \pg{210--245}.

\bibitem[Assi {\em et~al.\/}(2006)Assi, Meneghini, Aranha, Bearman \&
  Casaprima]{assi2006experimental}
{\sc \au{Assi, GR da~S}, \au{Meneghini, Julio~Romano}, \au{Aranha, Jos{\'e}
  Augusto~Penteado}, \au{Bearman, Peter~W} \& \au{Casaprima, Enrique}}
  \yr{2006}  \at{Experimental investigation of flow-induced vibration
  interference between two circular cylinders}.  \jt{Journal of fluids and
  structures}  \bvol{22}~(6-7),  \pg{819--827}.

\bibitem[Assi {\em et~al.\/}(2010)Assi, Bearman \& Meneghini]{assi2010wake}
{\sc \au{Assi, Gustavo~RS}, \au{Bearman, PW} \& \au{Meneghini, JR}} \yr{2010}
  \at{On the wake-induced vibration of tandem circular cylinders: the vortex
  interaction excitation mechanism}.  \jt{Journal of Fluid Mechanics}
  \bvol{661},  \pg{365--401}.

\bibitem[Bekemeyer \& Timme(2019)]{bekemeyer2019flexible}
{\sc \au{Bekemeyer, P} \& \au{Timme, S}} \yr{2019}  \at{Flexible aircraft gust
  encounter simulation using subspace projection model reduction}.
  \jt{Aerospace Science and Technology}  \bvol{86},  \pg{805--817}.

\bibitem[Bernitsas(2016)]{bernitsas2016harvesting}
{\sc \au{Bernitsas, Michael~M}} \yr{2016}  \at{Harvesting energy by flow
  included motions}.  \jt{Springer handbook of ocean engineering}  \pg{pp.
  1163--1244}.

\bibitem[Bernitsas {\em et~al.\/}(2008)Bernitsas, Raghavan, Ben-Simon \&
  Garcia]{bernitsas2008vivace}
{\sc \au{Bernitsas, Michael~M}, \au{Raghavan, Kamaldev}, \au{Ben-Simon, Y} \&
  \au{Garcia, EMH}} \yr{2008}  \at{Vivace (vortex induced vibration aquatic
  clean energy): A new concept in generation of clean and renewable energy from
  fluid flow}.  \jt{Journal of offshore mechanics and Arctic engineering}
  \bvol{130}~(4).

\bibitem[Bokaian \& Geoola(1984)]{bokaian1984wake}
{\sc \au{Bokaian, A} \& \au{Geoola, F}} \yr{1984}  \at{Wake-induced galloping
  of two interfering circular cylinders}.  \jt{Journal of Fluid Mechanics}
  \bvol{146},  \pg{383--415}.

\bibitem[Borazjani \& Sotiropoulos(2009)]{borazjani2009vortex}
{\sc \au{Borazjani, Iman} \& \au{Sotiropoulos, Fotis}} \yr{2009}
  \at{Vortex-induced vibrations of two cylinders in tandem arrangement in the
  proximity--wake interference region}.  \jt{Journal of fluid mechanics}
  \bvol{621},  \pg{321--364}.

\bibitem[Brika \& Laneville(1999)]{brika1999flow}
{\sc \au{Brika, D} \& \au{Laneville, A}} \yr{1999}  \at{The flow interaction
  between a stationary cylinder and a downstream flexible cylinder}.
  \jt{Journal of Fluids and Structures}  \bvol{13}~(5),  \pg{579--606}.

\bibitem[Carini {\em et~al.\/}(2014)Carini, Giannetti \&
  Auteri]{carini2014first}
{\sc \au{Carini, Marco}, \au{Giannetti, Flavio} \& \au{Auteri, Franco}}
  \yr{2014}  \at{First instability and structural sensitivity of the flow past
  two side-by-side cylinders}.  \jt{Journal of fluid mechanics}  \bvol{749},
  \pg{627--648}.

\bibitem[Chai {\em et~al.\/}(2021)Chai, Gao, Ankay, Li \&
  Zhang]{chai2021aeroelastic}
{\sc \au{Chai, Yuyang}, \au{Gao, Wei}, \au{Ankay, Benjamin}, \au{Li, Fengming}
  \& \au{Zhang, Chuanzeng}} \yr{2021}  \at{Aeroelastic analysis and flutter
  control of wings and panels: a review}.  \jt{International Journal of
  Mechanical System Dynamics}  \bvol{1}~(1),  \pg{5--34}.

\bibitem[Chen {\em et~al.\/}(2020)Chen, Ji, Alam, Williams \&
  Xu]{chen2020numerical}
{\sc \au{Chen, Weilin}, \au{Ji, Chunning}, \au{Alam, Md~Mahbub}, \au{Williams,
  John} \& \au{Xu, Dong}} \yr{2020}  \at{Numerical simulations of flow past
  three circular cylinders in equilateral-triangular arrangements}.
  \jt{Journal of Fluid Mechanics}  \bvol{891},  \pg{A14}.

\bibitem[Chen {\em et~al.\/}(2018)Chen, Ji, Williams, Xu, Yang \&
  Cui]{chen2018vortex}
{\sc \au{Chen, Weilin}, \au{Ji, Chunning}, \au{Williams, John}, \au{Xu, Dong},
  \au{Yang, Lihong} \& \au{Cui, Yuting}} \yr{2018}  \at{Vortex-induced
  vibrations of three tandem cylinders in laminar cross-flow: Vibration
  response and galloping mechanism}.  \jt{Journal of Fluids and Structures}
  \bvol{78},  \pg{215--238}.

\bibitem[Conciauro \& Puglisi(1981)]{conciauro1981meaning}
{\sc \au{Conciauro, G} \& \au{Puglisi, M}} \yr{1981}  \at{Meaning of the
  negative impedance}.  \jt{NASA STI/Recon Technical Report N}  \bvol{82},
  \pg{14458}.

\bibitem[Corrochano {\em et~al.\/}(2023)Corrochano, Sierra-Aus{\'\i}n, Martin,
  Fabre \& Le~Clainche]{corrochano2023mode}
{\sc \au{Corrochano, Adri{\'a}n}, \au{Sierra-Aus{\'\i}n, Javier}, \au{Martin,
  JA}, \au{Fabre, David} \& \au{Le~Clainche, Soledad}} \yr{2023}  \at{Mode
  selection in concentric jets: the steady--steady 1: 2 resonant mode
  interaction with o (2) symmetry}.  \jt{Journal of Fluid Mechanics}
  \bvol{971},  \pg{A30}.

\bibitem[Fabre {\em et~al.\/}(2018)Fabre, Citro, Ferreira~Sabino, Bonnefis,
  Sierra, Giannetti \& Pigou]{fabre2018practical}
{\sc \au{Fabre, David}, \au{Citro, Vincenzo}, \au{Ferreira~Sabino, D},
  \au{Bonnefis, Paul}, \au{Sierra, Javier}, \au{Giannetti, Flavio} \&
  \au{Pigou, Maxime}} \yr{2018}  \at{A practical review on linear and nonlinear
  global approaches to flow instabilities}.  \jt{Applied Mechanics Reviews}
  \bvol{70}~(6).

\bibitem[Fabre {\em et~al.\/}(2019)Fabre, Longobardi, Bonnefis \&
  Luchini]{fabre2019acoustic}
{\sc \au{Fabre, David}, \au{Longobardi, Raffaele}, \au{Bonnefis, Paul} \&
  \au{Luchini, Paolo}} \yr{2019}  \at{The acoustic impedance of a laminar
  viscous jet through a thin circular aperture}.  \jt{Journal of Fluid
  Mechanics}  \bvol{864},  \pg{5--44}.

\bibitem[Fabre {\em et~al.\/}(2020)Fabre, Longobardi, Citro \&
  Luchini]{fabre2020acoustic}
{\sc \au{Fabre, David}, \au{Longobardi, Raffaele}, \au{Citro, Vincenzo} \&
  \au{Luchini, Paolo}} \yr{2020}  \at{Acoustic impedance and hydrodynamic
  instability of the flow through a circular aperture in a thick plate}.
  \jt{Journal of Fluid Mechanics}  \bvol{885},  \pg{A11}.

\bibitem[Griffin \& Ramberg(1982)]{griffin1982some}
{\sc \au{Griffin, OM} \& \au{Ramberg, SE}} \yr{1982}  \at{Some recent studies
  of vortex shedding with application to marine tubulars and risers}.
  \jt{Journal of Energy Resources Technology}  \bvol{104}~(1),  \pg{2--13}.

\bibitem[Griffith {\em et~al.\/}(2017)Griffith, Jacono, Sheridan \&
  Leontini]{griffith2017flow}
{\sc \au{Griffith, Martin~D}, \au{Jacono, David~Lo}, \au{Sheridan, John} \&
  \au{Leontini, Justin~S}} \yr{2017}  \at{Flow-induced vibration of two
  cylinders in tandem and staggered arrangements}.  \jt{Journal of Fluid
  Mechanics}  \bvol{833},  \pg{98--130}.

\bibitem[Hassig(1971)]{hassig1971approximate}
{\sc \au{Hassig, Hermann~J}} \yr{1971}  \at{An approximate true damping
  solution of the flutter equation by determinant iteration.}  \jt{Journal of
  Aircraft}  \bvol{8}~(11),  \pg{885--889}.

\bibitem[Hecht(2012)]{hecht2012new}
{\sc \au{Hecht, Fr{\'e}d{\'e}ric}} \yr{2012}  \at{New development in
  freefem++}.  \jt{Journal of numerical mathematics}  \bvol{20}~(3-4),
  \pg{251--266}.

\bibitem[Hosseini {\em et~al.\/}(2021)Hosseini, Griffith \&
  Leontini]{hosseini2021flow}
{\sc \au{Hosseini, Negar}, \au{Griffith, Martin~D} \& \au{Leontini, Justin~S}}
  \yr{2021}  \at{Flow states and transitions in flows past arrays of tandem
  cylinders}.  \jt{Journal of Fluid Mechanics}  \bvol{910},  \pg{A34}.

\bibitem[Houtman \& Timme(2023)]{houtman2023global}
{\sc \au{Houtman, Jelle} \& \au{Timme, Sebastian}} \yr{2023}  \at{Global
  stability analysis of elastic aircraft in edge-of-the-envelope flow}.
  \jt{Journal of Fluid Mechanics}  \bvol{967},  \pg{A4}.

\bibitem[Huera-Huarte \& Gharib(2011)]{huera2011vortex}
{\sc \au{Huera-Huarte, FJ} \& \au{Gharib, M}} \yr{2011}  \at{Vortex-and
  wake-induced vibrations of a tandem arrangement of two flexible circular
  cylinders with far wake interference}.  \jt{Journal of Fluids and Structures}
   \bvol{27}~(5-6),  \pg{824--828}.

\bibitem[Kim {\em et~al.\/}(2009{\natexlab{{\em a\/}}})Kim, Alam, Sakamoto \&
  Zhou]{kim2009bflow}
{\sc \au{Kim, Sangil}, \au{Alam, Md~Mahbub}, \au{Sakamoto, Hiroshi} \&
  \au{Zhou, Yu}} \yr{2009{\natexlab{{\em a\/}}}}  \at{Flow-induced vibration of
  two circular cylinders in tandem arrangement. part 2: Suppression of
  vibrations}.  \jt{Journal of wind engineering and industrial aerodynamics}
  \bvol{97}~(5-6),  \pg{312--319}.

\bibitem[Kim {\em et~al.\/}(2009{\natexlab{{\em b\/}}})Kim, Alam, Sakamoto \&
  Zhou]{kim2009aflow}
{\sc \au{Kim, Sangil}, \au{Alam, Md~Mahbub}, \au{Sakamoto, Hiroshi} \&
  \au{Zhou, Yu}} \yr{2009{\natexlab{{\em b\/}}}}  \at{Flow-induced vibrations
  of two circular cylinders in tandem arrangement. part 1: Characteristics of
  vibration}.  \jt{Journal of Wind Engineering and Industrial Aerodynamics}
  \bvol{97}~(5-6),  \pg{304--311}.

\bibitem[King \& Johns(1976)]{king1976wake}
{\sc \au{King, R} \& \au{Johns, DJ}} \yr{1976}  \at{Wake interaction
  experiments with two flexible circular cylinders in flowing water}.
  \jt{Journal of Sound and Vibration}  \bvol{45}~(2),  \pg{259--283}.

\bibitem[Lee {\em et~al.\/}(2019)Lee, Qi, Zhou \& Lua]{lee2019vortex}
{\sc \au{Lee, Yin~Jen}, \au{Qi, Yi}, \au{Zhou, Guangya} \& \au{Lua, Kim~Boon}}
  \yr{2019}  \at{Vortex-induced vibration wind energy harvesting by
  piezoelectric mems device in formation}.  \jt{Scientific reports}
  \bvol{9}~(1),  \pg{20404}.

\bibitem[Mittal \& Kumar(2001)]{mittal2001flow}
{\sc \au{Mittal, S} \& \au{Kumar, Vinod}} \yr{2001}  \at{Flow-induced
  oscillations of two cylinders in tandem and staggered arrangements}.
  \jt{Journal of Fluids and Structures}  \bvol{15}~(5),  \pg{717--736}.

\bibitem[Navrose \& Mittal(2016)]{navrose2016lock}
{\sc \au{Navrose} \& \au{Mittal, Sanjay}} \yr{2016}  \at{Lock-in in
  vortex-induced vibration}.  \jt{Journal of Fluid Mechanics}  \bvol{794},
  \pg{565--594}.

\bibitem[Papaioannou {\em et~al.\/}(2008)Papaioannou, Yue, Triantafyllou \&
  Karniadakis]{papaioannou2008effect}
{\sc \au{Papaioannou, GV}, \au{Yue, DKP}, \au{Triantafyllou, MS} \&
  \au{Karniadakis, GE}} \yr{2008}  \at{On the effect of spacing on the
  vortex-induced vibrations of two tandem cylinders}.  \jt{Journal of Fluids
  and Structures}  \bvol{24}~(6),  \pg{833--854}.

\bibitem[Patel {\em et~al.\/}(2025)Patel, Sun, Yang \& Zhu]{patel2025coupled}
{\sc \au{Patel, Kuntal}, \au{Sun, Jun}, \au{Yang, Zixuan} \& \au{Zhu, Xiaojue}}
  \yr{2025}  \at{Coupled liquid--gas flow over a submerged cylinder: interface
  topology, wake structure and hydrodynamic lift}.  \jt{Journal of Fluid
  Mechanics}  \bvol{1008},  \pg{A10}.

\bibitem[Pfister {\em et~al.\/}(2019)Pfister, Marquet \&
  Carini]{pfister2019linear}
{\sc \au{Pfister, Jean-Lou}, \au{Marquet, Olivier} \& \au{Carini, Marco}}
  \yr{2019}  \at{Linear stability analysis of strongly coupled fluid--structure
  problems with the arbitrary-lagrangian--eulerian method}.  \jt{Computer
  Methods in Applied Mechanics and Engineering}  \bvol{355},  \pg{663--689}.

\bibitem[Prasanth \& Mittal(2009{\natexlab{{\em a\/}}})]{prasanth2009flow}
{\sc \au{Prasanth, TK} \& \au{Mittal, S}} \yr{2009{\natexlab{{\em a\/}}}}
  \at{Flow-induced oscillation of two circular cylinders in tandem arrangement
  at low re}.  \jt{Journal of fluids and structures}  \bvol{25}~(6),
  \pg{1029--1048}.

\bibitem[Prasanth \& Mittal(2009{\natexlab{{\em b\/}}})]{prasanth2009vortex}
{\sc \au{Prasanth, T.K.} \& \au{Mittal, Sanjay}} \yr{2009{\natexlab{{\em
  b\/}}}}  \at{Vortex-induced vibration of two circular cylinders at low
  reynolds number}.  \jt{Journal of Fluids and Structures}  \bvol{25}~(4),
  \pg{731--741}.

\bibitem[Sabino {\em et~al.\/}(2020)Sabino, Fabre, Leontini \&
  Jacono]{sabino2020vortex}
{\sc \au{Sabino, Diogo}, \au{Fabre, David}, \au{Leontini, JS} \& \au{Jacono,
  D~Lo}} \yr{2020}  \at{Vortex-induced vibration prediction via an impedance
  criterion}.  \jt{Journal of Fluid Mechanics}  \bvol{890}.

\bibitem[Sierra {\em et~al.\/}(2020)Sierra, Fabre \&
  Citro]{sierra2020efficient}
{\sc \au{Sierra, Javier}, \au{Fabre, David} \& \au{Citro, Vincenzo}} \yr{2020}
  \at{Efficient stability analysis of fluid flows using complex mapping
  techniques}.  \jt{Computer Physics Communications}  \bvol{251},  \pg{107100}.

\bibitem[Sierra-Ausin {\em et~al.\/}(2022)Sierra-Ausin, Citro \&
  Giannetti]{sierra2022acoustic}
{\sc \au{Sierra-Ausin, J}, \au{Citro, V} \& \au{Giannetti, F}} \yr{2022}
  \at{Acoustic instability prediction of the flow through a circular aperture
  in a thick plate via an impedance criterion}.  \jt{Journal of Fluid
  Mechanics}  \bvol{943},  \pg{A48}.

\bibitem[Sierra-Aus{\'\i}n {\em et~al.\/}(2022)Sierra-Aus{\'\i}n,
  Lorite-D{\'\i}ez, Jim{\'e}nez-Gonz{\'a}lez, Citro \&
  Fabre]{sierra2022unveiling}
{\sc \au{Sierra-Aus{\'\i}n, J}, \au{Lorite-D{\'\i}ez, M},
  \au{Jim{\'e}nez-Gonz{\'a}lez, JI}, \au{Citro, V} \& \au{Fabre, D}} \yr{2022}
  \at{Unveiling the competitive role of global modes in the pattern formation
  of rotating sphere flows}.  \jt{Journal of Fluid Mechanics}  \bvol{942},
  \pg{A54}.

\bibitem[Suzuki \& Inamuro(2011)]{suzuki2011effect}
{\sc \au{Suzuki, Kosuke} \& \au{Inamuro, Takaji}} \yr{2011}  \at{Effect of
  internal mass in the simulation of a moving body by the immersed boundary
  method}.  \jt{Computers \& Fluids}  \bvol{49}~(1),  \pg{173--187}.

\bibitem[Timme {\em et~al.\/}(2011)Timme, Marques \&
  Badcock]{timme2011transonic}
{\sc \au{Timme, S}, \au{Marques, Simao} \& \au{Badcock, KJ}} \yr{2011}
  \at{Transonic aeroelastic stability analysis using a kriging-based schur
  complement formulation}.  \jt{AIAA journal}  \bvol{49}~(6),  \pg{1202--1213}.

\bibitem[Tirri {\em et~al.\/}(2023)Tirri, Nitti, Sierra-Ausin, Giannetti \&
  de~Tullio]{tirri2023linear}
{\sc \au{Tirri, Antonia}, \au{Nitti, Alessandro}, \au{Sierra-Ausin, Javier},
  \au{Giannetti, Flavio} \& \au{de~Tullio, Marco~D}} \yr{2023}  \at{Linear
  stability analysis of fluid--structure interaction problems with an immersed
  boundary method}.  \jt{Journal of Fluids and Structures}  \bvol{117},
  \pg{103830}.

\bibitem[Williamson {\em et~al.\/}(2004)Williamson, Govardhan {\em
  et~al.\/}]{williamson2004vortex}
{\sc \au{Williamson, Charles~HK}, \au{Govardhan, R} \& \au{others}} \yr{2004}
  \at{Vortex-induced vibrations}.  \jt{Annual review of fluid mechanics}
  \bvol{36}~(1),  \pg{413--455}.

\bibitem[Zdravkovich(1987)]{zdravkovich1987effects}
{\sc \au{Zdravkovich, MM}} \yr{1987}  \at{The effects of interference between
  circular cylinders in cross flow}.  \jt{Journal of fluids and structures}
  \bvol{1}~(2),  \pg{239--261}.

\bibitem[Zhang {\em et~al.\/}(2024{\natexlab{{\em a\/}}})Zhang, Tang, Lu, Jin,
  An \& Cheng]{zhang2024flow}
{\sc \au{Zhang, Cheng}, \au{Tang, Guoqiang}, \au{Lu, Lin}, \au{Jin, Yan},
  \au{An, Hongwei} \& \au{Cheng, Liang}} \yr{2024{\natexlab{{\em a\/}}}}
  \at{Flow-induced vibration of two tandem square cylinders at low reynolds
  number: transitions among vortex-induced vibration, biased oscillation and
  galloping}.  \jt{Journal of Fluid Mechanics}  \bvol{986},  \pg{A10}.

\bibitem[Zhang {\em et~al.\/}(2024{\natexlab{{\em b\/}}})Zhang, Lu \&
  Zhang]{zhang2024global}
{\sc \au{Zhang, Zhiyu}, \au{Lu, Jianfeng} \& \au{Zhang, Xing}}
  \yr{2024{\natexlab{{\em b\/}}}}  \at{Global stability analysis of
  flow-induced-vibration problems using an immersed boundary method}.
  \jt{Journal of Fluids and Structures}  \bvol{130},  \pg{104187}.

\bibitem[Zhu {\em et~al.\/}(2024)Zhu, Zhong, Shao, Zhou \& Alam]{zhu2024fluid}
{\sc \au{Zhu, Hongjun}, \au{Zhong, Jiawen}, \au{Shao, Ze}, \au{Zhou, Tongming}
  \& \au{Alam, Md~Mahbub}} \yr{2024}  \at{Fluid-structure interaction among
  three tandem circular cylinders oscillating transversely at a low reynolds
  number of 150}.  \jt{Journal of Fluids and Structures}  \bvol{130},
  \pg{104204}.

\end{thebibliography}

\end{document}